\documentclass[12pt,preprint]{aastex}

\shorttitle{Photon Bubbles in Radiation-Dominated Disks}
\shortauthors{Turner et al.}

\newcommand{\divv}{{\bf\nabla\cdot v}}

\begin{document}
\title{The Effects of Photon Bubble Instability in Radiation-Dominated
Accretion Disks}

\author{N. J. Turner\altaffilmark{1,2}, O. M. Blaes\altaffilmark{1},
A. Socrates\altaffilmark{1,3,4}, M. C. Begelman\altaffilmark{5}, and
S. W. Davis\altaffilmark{1}}

\altaffiltext{1}{Physics Department, University of California, Santa
Barbara CA 93106}

\altaffiltext{2}{Jet Propulsion Laboratory MS 169-506, California
Institute of Technology, Pasadena CA 91109; neal.turner@jpl.nasa.gov}

\altaffiltext{3}{Department of Astrophysical Sciences, Princeton
University, Princeton NJ 08544}

\altaffiltext{4}{Hubble Fellow}

\altaffiltext{5}{JILA, University of Colorado, Boulder CO 80309}

\begin{abstract}
We examine the effects of photon bubble instability in
radiation-dominated accretion disks such as those found around black
holes in active galactic nuclei and X-ray binary star systems.  Two-
and three-dimensional numerical radiation MHD calculations of small
patches of disk are used.  Modes with wavelengths shorter than the gas
pressure scale height grow faster than the orbital frequency in the
disk surface layers.  The fastest growth rate observed is five times
the orbital frequency and occurs on nearly-vertical magnetic fields.
The spectrum of linear modes is in good agreement with a WKB analysis
that indicates still faster growth at unresolved scales, with a
maximum growth rate proportional to the gravitational acceleration and
inversely proportional to the gas sound speed.  Disturbances reaching
non-linear amplitudes steepen into trains of shocks similar to a 1-D
periodic non-linear analytic solution.  Variations in propagation
speed result in merging of adjacent fronts, and over time the shock
spacing and amplitude increase.  Growth is limited by the strength of
the magnetic field.  The shock train structure is disrupted when the
ram pressure of the disturbances exceeds the magnetic pressure.  The
maximum horizontal density variations are comparable to the ratio of
magnetic to gas pressure, and in our calculations exceed one hundred.
Under the conditions considered, radiation diffuses through the
inhomogeneneous flow five times faster than through the initial
hydrostatic equilibrium, and the net cooling rate is several times
greater than in a similar calculation without magnetic fields that
shows the effects of convection.  These results indicate that photon
bubbles may be important in cooling radiation-dominated accretion
disks.  The Shaviv type~I global instability grows faster than the
orbital frequency in calculations of the disk surface layers with
lower boundaries of fixed temperature, but is weak or absent in
calculations spanning the disk thickness.
\end{abstract}


%%%%%%%%%%%%%%%%%%%%%%%%%%%%%%%%%%%%%%%%%%%%%%%%%%%%%%%%%%%%%%%%%%%%%%%%%%%%%%%
\section{INTRODUCTION\label{sec:intro}}

Black hole systems with luminosities between about 1\% and 100\% of
the Eddington limit are thought to be powered by accretion through a
geometrically thin disk supported by rotation.  Near the hole, the
internal radiation pressure greatly exceeds the gas pressure and
determines the thickness of the disk.  In the standard picture,
angular momentum is transferred outward within the disk by a torque of
unspecified origin, proportional to the vertically-averaged gas plus
radiation pressure.  The released gravitational energy escapes by
diffusion of photons to the disk faces \citep{ss73}.  The model is
unstable to perturbations in the mass flow \citep{le74} and heating
rates \citep{ss76}, and no steady accretion is possible in
radiation-pressure dominated regions in the standard picture.
However, the thermal instability is absent if additional cooling
processes operate and the cooling rate increases with disk thickness
faster than the heating rate \citep{pi78}.

The evolution of disks is governed by torques together with heating
and cooling processes.  In the inner parts of black hole disks, the
accretion stresses are due to magnetic forces.  Magneto-rotational
instability (MRI) leads to turbulence in which magnetic fields linking
material at different distances from the hole transfer angular
momentum outward \citep{bh91,bh98}.  The gas is heated by the
dissipation of the magnetic fields and the turbulence through
microscopic resistivity and viscosity.  In radiation-dominated disks,
kinetic energy is also converted directly to photon energy by
radiative diffusion damping of compressive motions \citep{ak98}.  The
strength of the magnetic fields is regulated by generation through MRI
and losses through buoyancy \citep{sr84,sc89} and dissipation.  In a
vertically-stratified 3-D local shearing-box radiation-MHD
calculation, the magnetic pressure is less than the gas pressure near
the midplane, and greater than the gas but less than the radiation
pressure in surface layers \citep{tu04}.  Field lines close to the
midplane are tangled while those near the disk surface lie mostly
along the direction of orbital motion.

\begin{figure}
\epsscale{0.59}
\plotone{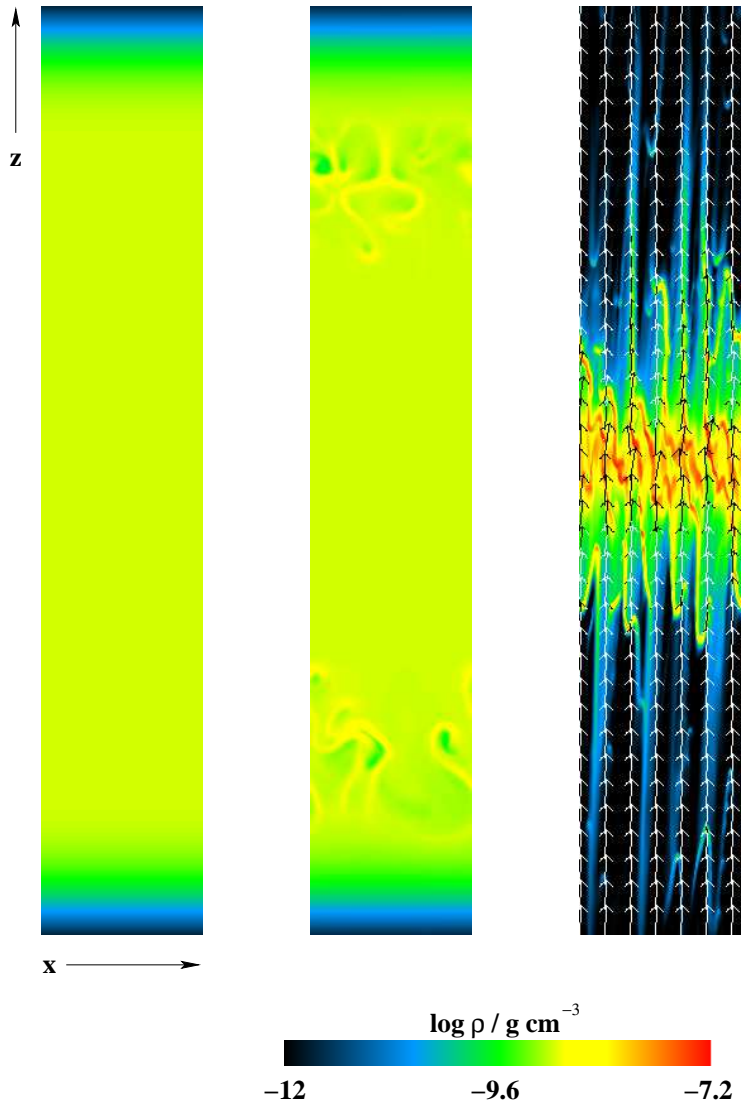}
\caption{Snapshots of the gas density at 1.5~orbits in three
calculations of a small patch of accretion disk, centered 20
Schwarzschild radii from a black hole of $10^8$ M$_\odot$.  Neither
viscous dissipation nor differential rotation is included.  The domain
extends 1.15~Shakura-Sunyaev scale-heights $H$ above and below the
midplane and the width is $0.4 H$.  At left are results from a
one-dimensional calculation in which the gas is cooled by vertical
radiation diffusion and contracts slightly toward the midplane.  At
center in a two-dimensional calculation, convection starts in the
outer layers.  The version shown at right differs only in including a
magnetic field, indicated by arrows, the longest corresponding to
6000~Gauss.  The field is initially uniform, inclined $87^\circ$ from
horizontal, with pressure 10\% of the midplane radiation pressure.
Evolution is rapid due to photon bubble instability.  The figure is
also available as three MPEG animations showing the top half of the
domain.
\label{fig:3up}}
\end{figure}

\begin{figure}
\epsscale{0.7}
\plotone{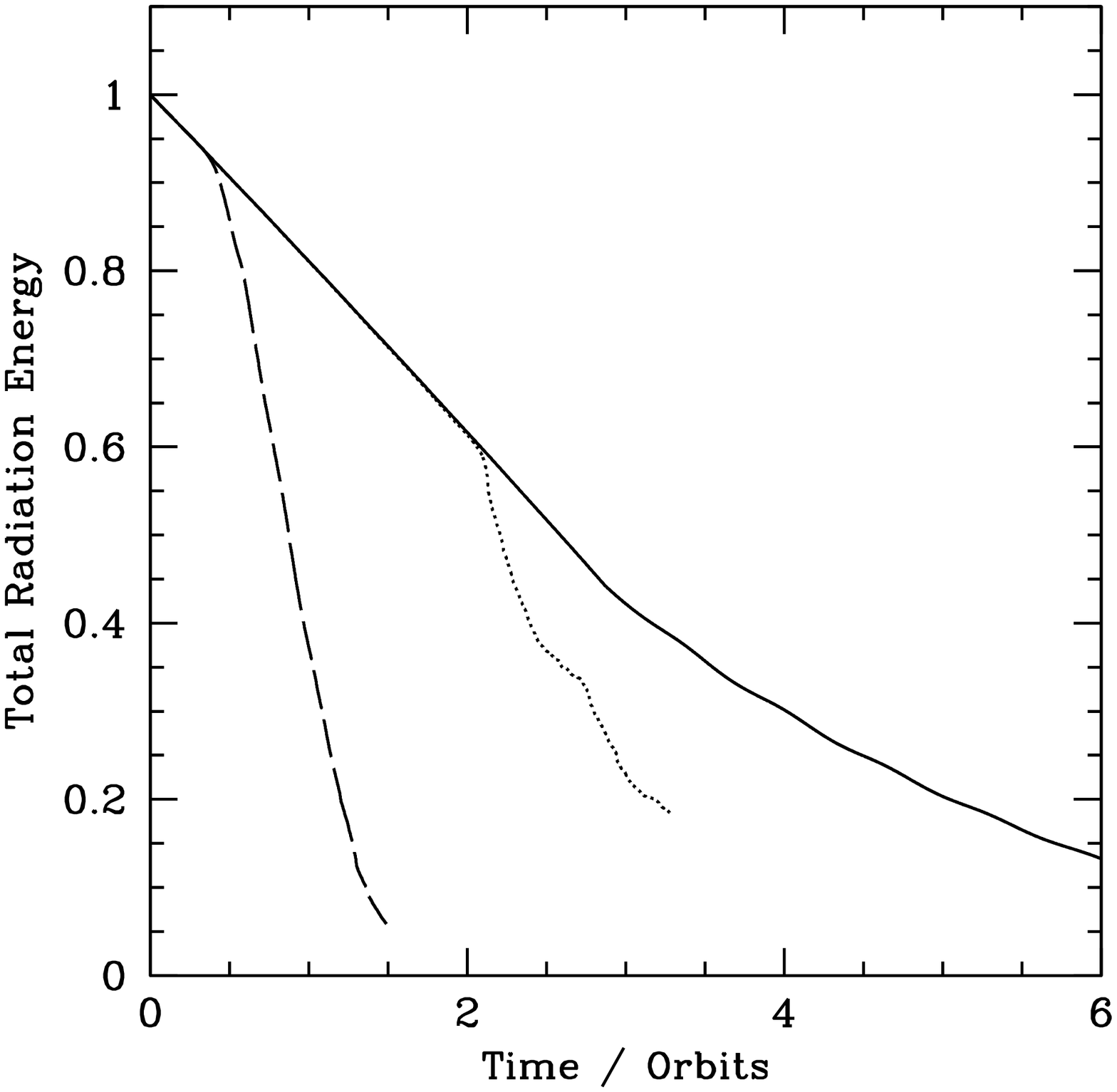}
\caption{Total radiation energy versus time in the three calculations
shown in figure~\ref{fig:3up}.  The diffusion calculation is indicated
by the solid curve, the convection calculation by the dotted curve,
and the photon bubble calculation by the dashed curve.  Energies are
normalized to the initial value.  The radiation energy falls by half
after 2.6~orbits in the diffusion case, 2.2~orbits in the convection
case and 0.9~orbits in the photon bubble case.
\label{fig:cool}}
\end{figure}

Steady accretion through a thin disk requires that local cooling
balance the vertically-integrated heating.  In this article we focus
on how radiation-dominated disks cool.  Several dynamical
instabilities have linear growth rates similar to or faster than the
orbital frequency, and may lead to vertical energy transport in
turbulence driven by the MRI.  The instabilities considered are
convection, the Shaviv modes and photon bubbles and have the following
properties.
\begin{enumerate}
\item The standard Shakura-Sunyaev model is convectively unstable
\citep{bkb77}.  Convection in two-dimensional radiation hydrodynamical
disk calculations carries energy vertically at a rate similar to
diffusion.  The extra losses are not sufficient to quench the thermal
instability \citep{ak01}.  Convection may be absent if the heating
from dissipation of magnetic fields is concentrated at low column
depths due to magnetic buoyancy \citep{sz94,tu04}.
\item Radiation-supported atmospheres through which photons diffuse in
less than a sound-crossing time may be subject to a global linear
instability even when convectively stable \citep{sh01}.  The resulting
overturning motions might lead to formation of low-density chimneys
where the radiative flux is enhanced.
\item Displacements of gas along magnetic field lines can be
overstable, leading to growing, propagating density variations known
as photon bubbles \citep{ar92,ga98}.  Growth is fastest at wavelengths
shorter than the gas pressure scale height, and in the
short-wavelength limit the instability is due to radiative driving of
the density fluctuations found in slow magneto\-sonic waves
\citep{bs01}.  Numerical results on neutron star accretion columns
indicate that short-wavelength photon bubble modes saturate at small
amplitudes, and the later evolution of the instability is dominated by
longer wavelengths \citep{ha97}.  Photon bubble instability may lead
to the development of trains of propagating shocks.  The flux of
radiation diffusing through the low-density gaps between the shocks
could be substantially greater than in the hydrostatic atmosphere
\citep{be01}.
\end{enumerate}
The effects of these three dynamical instabilities are explored by
numerically solving the equations of radiation MHD in a patch of disk
centered 20~Schwarzschild radii $R_S=2GM/c^2$ from a black hole of
mass $M=10^8$ M$_\odot$.  The rapid cooling possible through
convection and photon bubble instability is illustrated in
figures~\ref{fig:3up} and~\ref{fig:cool}.  The domain height here is
2.3~times the disk thickness and the width is 17\% of the height.  The
initial state is a standard Shakura-Sunyaev model with accretion rate
10\% of the Eddington limit for a 10\% radiative efficiency,
constructed using a ratio of height-averaged accretion stress to gas
plus radiation pressure $\alpha=0.06$.  The dissipation rate per unit
volume is assumed proportional to the density at each height.  During
the calculations, the $\alpha$-viscosity is omitted and there is no
further injection of energy.  We measure the rate at which the initial
reservoir of energy is depleted by loss of radiation through the disk
surfaces.  To study the cooling processes separately from the energy
release resulting from the MRI, we neglect the radial gradient in
angular velocity.  The side boundaries are periodic, and the top and
bottom boundaries allow gas, radiation and magnetic fields to escape.
The domain is divided into $128\times 736$ zones, and the initial
equilibrium is disturbed slightly by applying random density
perturbations with probability uniformly distributed between $-1$\%
and $+1$\%.  Any effects of the boundaries are reduced by applying the
perturbations only in the disk interior where the density is greater
than half the midplane value.  In the case with a magnetic field,
photon bubble instability leads to large density inhomogeneities.
Radiation escapes through the patchy atmosphere in this case five
times faster than through the smooth density distribution in the
calculation with vertical diffusion alone.

The equations solved and numerical methods are described in
section~\ref{sec:method}, and the construction of the initial
equilibrium in section~\ref{sec:ic}.  In sections~\ref{sec:convection}
and~\ref{sec:shavivi} we consider the effects of convection and the
\cite{sh01} instabilities, which require no magnetic fields.  The
growth rates of linear radiation MHD disturbances are compared against
the results of the \cite{bs03} WKB analysis in
section~\ref{sec:linear}.  The linear photon bubble modes develop into
trains of shock fronts.  The shock structure is compared against the
\cite{be01} non-linear analytic solution in section~\ref{sec:shocks}
and limits on the growth of the shock trains are discussed.  A summary
and conclusions are in section~\ref{sec:conc}.

%%%%%%%%%%%%%%%%%%%%%%%%%%%%%%%%%%%%%%%%%%%%%%%%%%%%%%%%%%%%%%%%%%%%%%%%%%%%%%%
\section{EQUATIONS AND METHOD OF SOLUTION\label{sec:method}}

The radiation magneto\-hydrodynamics (MHD) equations solved are
frequency-averaged and include only terms that are of order unity in
$v/c$ in at least one optical depth regime.  A Cartesian coordinate
system is used, with $x$- and $y$-axes lying in the disk midplane and
$z$-axis vertical.  Ideal MHD is assumed.  Conservation of mass, gas
momentum, radiation momentum, gas energy, and radiation energy, and
the evolution of magnetic fields are described by
\begin{equation}\label{eqn:cty}
{{D\rho}\over{D t}}+\rho\divv=0,
\end{equation}
\begin{equation}\label{eqn:eomg}
\rho{{D{\bf v}}\over{D t}} = -{\bf\nabla}p
	+ {1\over 4\pi}({\bf\nabla\times B}){\bf\times B}
        + {\chi_F\rho\over c}{\bf F} - \rho\Omega^2 z {\bf\hat z},
\end{equation}
\begin{equation}\label{eqn:eomr}
{\bf F} = -{c\Lambda\over\chi_F\rho}{\bf\nabla}E,
\end{equation}
\begin{equation}\label{eqn:eoeg}
\rho{D\over D t}\left({e\over\rho}\right) =
	- p\divv - \kappa_P\rho(4\pi B - c E),
\end{equation}
\begin{equation}\label{eqn:eoer}
\rho{D\over D t}\left({E\over\rho}\right) =
	- {\bf\nabla\cdot F} - {\bf\nabla v}:\mathrm{P}
	+ \kappa_P\rho(4\pi B - c E),
\end{equation}
and
\begin{equation}\label{eqn:dbdt}
{\partial{\bf B}\over\partial t} = {\bf\nabla\times}({\bf v\times B})
\end{equation}
\citep{mm84,sm92,bs03}.  The vertical component of the gravity of the
black hole is included by an acceleration $g=\Omega^2 z$, proportional
to the square of the Keplerian orbital frequency $\Omega =
\left(GM/R^3\right)^{1/2}$ at domain center.  The center is placed at
radius $R = 20 R_S$ and differential orbital motion is neglected.  The
opacities included are the flux-mean total opacity $\chi_F$, which is
dominated by electron scattering, and the Planck-mean free-free
absorption opacity $\kappa_P$.  The radiation is assumed to be
sufficiently similar to a blackbody at the gas temperature that
differences between the Planck-weighted and intensity-weighted
frequency averages of the absorption opacity can be neglected.  The
set of equations~\ref{eqn:cty} to~\ref{eqn:dbdt} is closed using an
ideal gas equation of state $p=(\gamma-1)e$, with $\gamma=5/3$, and an
Eddington tensor $\rm f$ relating the radiation pressure tensor to the
radiation energy density through ${\rm P}={\rm f} E$.  The angular
dependence of the radiation field which determines the Eddington
tensor is treated approximately using the flux-limited diffusion (FLD)
method.  The Eddington factor and the flux limiter $\Lambda$ approach
$1/3$ in optically-thick regions.  The limiter is reduced in
optically-thin regions according to the prescription of \cite{lp81}
equation~22, ensuring that photons transport energy no faster than
light speed.

The equations are integrated using the Zeus MHD code
\citep{sn92a,sn92b} with its FLD module \citep{ts01}.  We make two
kinds of calculation.  Those spanning the thickness of the disk as in
figure~\ref{fig:3up} show the overall cooling effects of the
instabilities, while calculations of small patches of the surface
layers are used for higher-resolution studies that are compared with
analytic results.  The timestep in the surface-layer calculations is
limited by the diffusion step $\Delta t_D = (\Delta x)^2/(N D_{\rm
max})$.  The diffusion step is approximately the time for radiation to
diffuse across the lowest-density grid zone, and depends on the grid
spacing $\Delta x$, number of spatial dimensions $N$, and
domain-maximum radiation diffusion coefficient
$D=c\Lambda/(\chi_F\rho)$.  The numerical method is stable with longer
timesteps due to the implicit differencing scheme used for the
radiation source terms.  However in calculations with longer steps,
well-resolved short-wavelength linear photon bubble modes grow slower
than expected (section~\ref{sec:pbitau}).  Execution times depend on
the number of grid zones and the number of timesteps.  With the
timestep limited by diffusion, execution times are proportional to
$(\Delta x)^{-4}$ in 2-D and $(\Delta x)^{-5}$ in 3-D calculations, so
that high resolutions are obtained at considerable expense.

The domain for the calculations spanning the disk thickness is two
orders of magnitude larger than the surface-layer patches and the
diffusion timestep set by the initial minimum density is 30 times
shorter.  Results are obtained in a reasonable amount of computer time
by allowing timesteps up to $100 \Delta t_D$.  While accuracy is
sacrificed in the outermost layers, the timesteps are initially
shorter than the time for photons to diffuse across grid zones within
$1.04 H$ of the midplane, where the densities are more than 100 times
the minimum.  Extremely low densities and short timesteps are
prevented in the full-disk-thickness calculations using a density
floor.  Gas is added to the grid where needed to bring the density up
to 0.1\% of the initial value at the midplane.  The floor is not
reached until significant evolution has occurred, at 2.82~orbits in
the diffusion calculation, 2.05~orbits in the convection calculation
and 0.76~orbits in the photon bubble calculation.  In the calculations
of the surface layer patches, no floor is applied and densities are
allowed to become arbitrarily small.

%%%%%%%%%%%%%%%%%%%%%%%%%%%%%%%%%%%%%%%%%%%%%%%%%%%%%%%%%%%%%%%%%%%%%%%%%%%%%%%
\section{DOMAIN AND INITIAL CONDITIONS\label{sec:ic}}

\begin{figure}
\epsscale{0.60}
\plotone{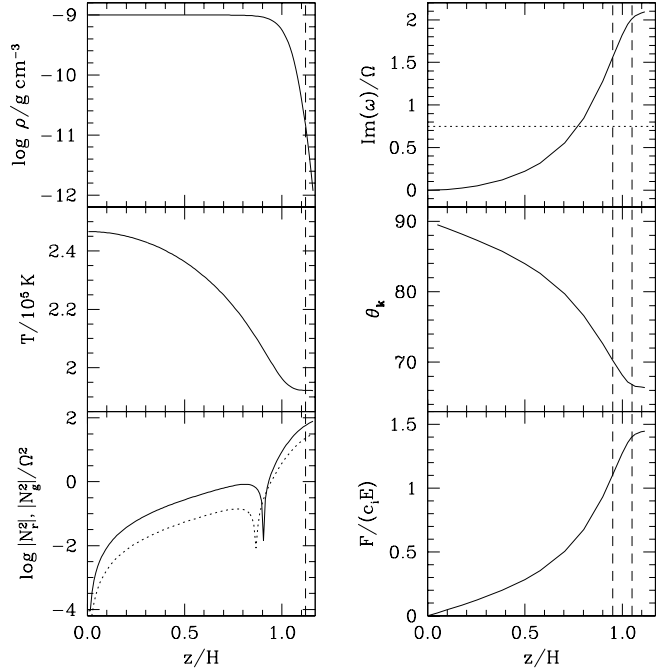}
\caption{Initial condition for the calculations of
figure~\ref{fig:3up}.  Only the top half of the domain is shown.
Density is in the top left panel, temperature at middle left, and
squared radiation (solid) and gas (dotted) Brunt-V\"ais\"al\"a
frequencies at bottom left.  The squared B-V frequencies are negative
below about $0.9 H$, indicating convective instability, and positive
above, indicating stability.  Vertical dashed lines in the left column
indicate the photosphere.  Photon bubble instability of the initial
condition is illustrated in the right column for the case of a uniform
horizontal magnetic field with pressure 10\% of the midplane radiation
pressure.  The growth rate of the fastest photon bubble mode having
wavenumber equal to the inverse gas pressure scale height is plotted
at upper right, the angle between its wavevector and the horizontal at
middle right.  These are obtained by solving the dispersion relation
of \cite{bs03}, their equation~49.  Near the photosphere, the fastest
mode has growth rate more than twice the orbital frequency, and
wavefronts inclined $24^\circ$ from horizontal.  Close to the
midplane, growth is slow, and the fastest mode has wavefronts almost
parallel to the field.  An approximate criterion for photon bubble
instability here is that the flux exceed the product of radiation
energy density and isothermal gas sound speed \citep{bs03}.  The flux
ratio is shown at lower right.  A horizontal dotted line in the top
right panel indicates the maximum growth rate of the
magneto-rotational instability.  Vertical dashed lines in the right
column mark the boundaries of the calculations used to follow the
growth of short-wavelength modes in small patches of the surface
layers of the model disk.
\label{fig:ic}}
\end{figure}

The initial states for the calculations shown in figure~\ref{fig:3up}
are taken from a standard Shakura-Sunyaev model.  The structure is in
hydrostatic equilibrium and radiative balance, and heat is assumed to
be deposited at a rate proportional to the density at each height.
The construction of the initial condition is described below in
section~\ref{sec:hsefb}, and the initial state is shown in
figure~\ref{fig:ic}.  The disk interior is radiation-supported and the
half-thickness $H = \chi_F F_1 / (\Omega^2 c) = 0.46 R_S$ is set by
the surface radiation flux $F_1$.  Characteristic speeds are the
radiation sound speed $c_r=(\frac{4}{9}E/\rho)^{1/2}$ and isothermal
gas sound speed $c_i=(p/\rho)^{1/2}$.  Characteristic scales for
disturbances are obtained by matching the corresponding pressure
gradients to the gravity.  The radiation pressure scale height
$c_r^2/g$ is roughly equal to $H$ and the gas pressure scale height,
$c_i^2/g$, is shorter by $4/3$ times the ratio of radiation to gas
pressure.  The ratio of the pressures is~275 at the midplane so that
the ratio of the two pressure scale heights is~367.

The density in the disk interior is almost uniform.  The radiation
flux increases linearly with height and the radiation force balances
the vertical component of gravity.  The time for radiation to diffuse
from the midplane to a height $H$ is $3/(2\pi\alpha)$~orbits or eight
orbits.  In the disk surface layers, densities are lower, so the flux
varies little with height and gas pressure provides the extra support
needed for hydrostatic balance.  The thickness of the gas pressure
supported surface layer is approximately $c_i(z=H)/\Omega = 0.048 H$
and photons diffuse through this layer in about 0.01~orbit, much
shorter than the gas sound crossing time $1/\Omega = 0.16$~orbit.  Gas
and radiation reach thermal equilibrium through emission and
absorption of photons in a time $t_{eqm} = e/(c\kappa_P\rho E)$
ranging from $2\times 10^{-7}$~orbits at the midplane to $2\times
10^{-5}$~orbits at the lowest densities, so we choose to place the
entire domain initially in thermal equilibrium.  The total optical
depth through the disk is 9200.  The boundary conditions for the
full-disk-thickness calculations are discussed in
section~\ref{sec:bc}.  The initial and boundary conditions for the
calculations in small patches of the surface layers are described in
section~\ref{sec:icsurface}.

\subsection{Hydrostatic and Radiative Balance\label{sec:hsefb}}

The initial condition is constructed by vertically integrating the
equations of hydrostatic equilibrium and radiative balance, assuming
that the dissipation rate is proportional to the density, that gas and
radiation are in thermal equilibrium, and that the radiation diffusion
approximation holds.  The integration runs from midplane to outer
boundary.  The midplane density is chosen so that the surface density
matches the Shakura-Sunyaev model, and the midplane radiation pressure
is chosen so the vertically-integrated accretion stress matches the
assumed accretion rate $\dot M$.

The procedure in detail is as follows.  The radial structure of the
Shakura-Sunyaev model is used to determine the surface mass density,
\begin{equation}\label{eqn:sigma}
\Sigma = {8\sqrt{2}\over 3\chi_F\alpha} {L_E\over{{\dot M}c^2}}
\left({R\over R_S}\right)^{3/2} {\cal I}^{-1},
\end{equation}
and the flux through each face of the disk,
\begin{equation}\label{eqn:flux1}
F_1 = {3{\dot M}\Omega^2\over 8\pi} {\cal I}.
\end{equation}
The Eddington luminosity $L_E=4\pi cGM/\chi_F$ in
equation~\ref{eqn:sigma} is the value for which outward radiation
forces balance the gravity of the black hole in spherical symmetry.
The factor ${\cal I}=1-(r_{in}/r)^{1/2}$ occurs because the accretion
stress is assumed to approach zero at the disk inner edge, $r_{in}$,
placed at $3 R_S$.  The choice of conditions at the inner edge is
expected to have little effect on the outcome of the calculations at
$20 R_S$.  Given the surface density and flux, the vertical structure
is completely specified by the vertical component of the equation of
motion
\begin{equation}\label{eqn:hse}
-{d\over dz}\left(P+p\right) = \rho\Omega^2z
\end{equation}
and the summed gas and radiation energy equations
\begin{equation}\label{eqn:fb}
{dF\over dz} = {2\rho F_1\over\Sigma},
\end{equation}
together with boundary values for density and temperature.
Equations~\ref{eqn:hse} and~\ref{eqn:fb} are integrated simultaneously
from the midplane outward.  The midplane density $\rho_c$ is chosen so
that the surface density of the resulting structure is $\Sigma$.  The
surface density is measured between the top and bottom surfaces of the
disk at optical depth unity.  The midplane temperature is adjusted
until the height-integrated accretion stress is sufficient to produce
the accretion rate, so that
\begin{equation}
\int\alpha(P+p)dz = {{\dot M}\Omega\over 2\pi} {\cal I}.
\end{equation}
As with the surface density, the vertical integration extends to the
surfaces of unit optical depth.

The initial state is constructed assuming Eddington factors of
one-third throughout.  This is inaccurate above the photosphere, where
the specific intensity is expected to be greater looking down than up.
Direct integration of the transfer equation indicates that the
Eddington factor at zero optical depth is~0.42.  The smaller Eddington
factor used corresponds to shallower gradients in radiation energy
density and higher temperatures near the photosphere.  The higher
temperatures lead to slower growth of photon bubbles in the disk
surface layers, as the fastest growth rates are proportional to
$g/c_i$ \citep{bs03}.  Our calculations therefore place a lower bound
on the growth rates of photon bubbles in the Shakura-Sunyaev model.

\subsection{Boundary Conditions\label{sec:bc}}

Gas, magnetic fields, and radiation are allowed to flow out through
the upper and lower boundaries of the full-disk-thickness
calculations, and radiation is allowed also to diffuse out.  The
gradients of gas temperature and density across the boundaries are set
to zero.  The vertical velocity is restricted to zero or outward
values, so that gas may flow out, but not in.  The outflow boundary
condition on the magnetic field is imposed through the electromotive
force (EMF).  The gradient in the EMF is zero across the boundary
\citep{sn92b}.  Radiation energy densities outside the domain are
chosen so that the flux across the boundary is approximately equal to
the flux between the outermost pair of active zones.  The fluxes are
calculated using the diffusion coefficients obtained in the previous
timestep, so that the boundary condition corresponds exactly to
$\partial F_z/\partial z=0$ only when the flow is time-independent.
Should the flux be directed into the domain, the radiation energy
densities outside are chosen to make the flux zero, preventing
radiation from entering.  The side boundaries are periodic.

\subsection{Surface Layers\label{sec:icsurface}}

The domain for the calculations of the disk surface layers extends
from 0.95 to $1.05 H$ above the midplane.  The initial conditions for
these calculations are constructed assuming no dissipation occurs
within the domain, so the flux is independent of height.  The
resulting structure has density gradients slightly steeper than the
corresponding region in the calculations spanning the whole disk
thickness.  The temperature and density at the height in the full
vertical structure that corresponds to the center of the smaller
domain are used as starting conditions for the integration of the
hydrostatic equilibrium and flux balance equations, \ref{eqn:hse} and
\ref{eqn:fb}.  The flux throughout is set to the value at the height
corresponding to the bottom of the surface layer domain and the
dissipation term on the right-hand side of equation~\ref{eqn:fb} is
zero.  The vertical integration is first performed assuming that the
opacity is due to electron scattering alone.  A more accurate
equilibrium is obtained with a second integration, including free-free
opacities calculated using the temperatures and densities resulting
from the first approximation.  The resulting structure when placed in
a 1-D vertical radiation hydrodynamic calculation shows residual
motions initially less than $10^{-8}$ times the radiation acoustic
speed, and decreasing or constant over 10~orbits.  The structure is
convectively stable, with radiation and gas Brunt-V\"ais\"al\"a
frequencies ranging from 1.17 and $1.01\Omega$ at the lower boundary
to 7.19 and $4.55\Omega$ at the top.  The total optical depth is 287.
Characteristic lengths are the density scale height $\rho /
\left|\frac{d\rho}{dz}\right|$ and the gas pressure scale height
$c_i^2/g$.  The density scale height ranges from $0.236 H$ at the
bottom boundary to $0.0202 H$ at the top, and the density declines by
a factor 14 or $2.67$ $e$-foldings over the height of the surface
layer patch.  The gas pressure scale height is the longest distance
over which gas pressure disturbances can balance gravity.  It depends
on the temperature and gravity, which are almost uniform over the
domain.  The gas pressure scale height is $0.00250 H$ at domain
bottom, $0.00229$ at center and $0.00216$ at top.  The ratio of
radiation to gas pressure ranges from 121 at the lower boundary to
1500 at the top, and at domain center $z=H$ the ratio is 254.  At the
lower boundary in the surface layer calculations, the temperature and
density are fixed at their initial values, and the vertical velocity
is zero.  The fixed temperature means that the supply of radiation
energy diffusing into the patch from below is inexhaustible.  The side
and top boundaries are treated as in the calculations spanning the
disk thickness.

\section{CONVECTION\label{sec:convection}}

Radiation-dominated Shakura-Sunyaev equilibria are convectively
unstable \citep{bkb77}, with the fastest modes growing at about the
orbital frequency \citep{pk00}.  Greatest convective instability in
the initial condition for the calculations in figure~\ref{fig:3up}
occurs a little below the disk surface, at $0.8 H$, as shown in
figure~\ref{fig:ic}, lower left panel.  Growth begins in the numerical
calculations from initial random 1\% density perturbations applied
within $1.01 H$ of the midplane, where the density is greater than
half the midplane value.  Growth rates during the linear stage are
measured using Fourier power spectra of the horizontal velocities in a
square region spanning the domain width $0.4 H$ and extending from
$0.5$ to $0.9 H$ above the midplane.  The fastest mode has horizontal
and vertical wavelengths 0.2 and $0.4 H$ and grows at $0.72 \Omega$.
The fastest mode in the corresponding region below the midplane has
horizontal and vertical wavelengths 0.13 and $0.4 H$ and grows at
$0.77 \Omega$.  These fastest modes are well-resolved, with 40 or more
grid zones per wavelength.  Among the solutions of the local linear
dispersion relation, \cite{bs03} equation~49, the same two wavevectors
grow fastest.  The analytic growth rates of both modes peak at height
$0.80 H$, where the rates are approximately equal and are $0.76
\Omega$.  At the upper boundary of the measurement region at $0.9 H$
there is convective stability, while at the lower boundary at $0.5 H$,
the dispersion relation indicates both modes grow at $0.43 \Omega$.
The differences between the fastest modes in the numerical and
analytic solutions are much less than the range in local analytic
growth rates across the region.  The WKB approximation is marginally
applicable to the fastest modes, as the radiation pressure at the top
of the region is 1.7 times less than at the bottom.  Nevertheless the
WKB solutions and the numerical results are consistent.

Once the disturbances reach non-linear amplitudes in the convection
calculation, overdense gas falls toward the midplane and underdense
material rises toward the vertical boundaries.  Radiation is carried
with the rising gas, leading to an overall increase in the cooling
rate (figure~\ref{fig:cool}).  The mean cooling rate between 2.25 and
3.25~orbits is 1.6~times that in the diffusion calculation.  The
horizontally-averaged ram pressure of the motions is greater than the
gas pressure and less than the radiation pressure.  The horizontal
separation between dense sinking plumes of gas is about equal to the
domain width, so we check whether cooling is limited by the width
using a version with the horizontal size doubled and the number of
zones in the horizontal direction increased to 256.  The cooling rate
is similar and the radiation energy falls to half its initial value in
2.1~orbits.

\section{OVERTURNING MODES\label{sec:shavivi}}

Results of a linear stability analysis indicate that unmagnetized
electron-scattering atmospheres transmitting a flux greater than about
half the local Eddington limit are subject to two instabilities driven
by the rapid diffusion of radiation with respect to the gas
\citep{sh01}.  The analysis is 2-D, and is local in the horizontal
direction and global in the vertical direction.  The temperature and
gravity used by Shaviv are appropriate to a white dwarf envelope
during a nova outburst.  The first of the modes, type~I, is
stationary, with density and radiation pressure disturbances
anticorrelated, and was seen only in calculations with
fixed-temperature lower boundary.  Its growth time is similar to the
time for gas sound waves to cross the density scale height.  The
second mode, type~II, is propagating, with a phase lag between density
and radiation pressure disturbances, and appears in calculations with
a variety of boundary conditions.  Its growth is an order of magnitude
slower.

\begin{figure}
\epsscale{1}
\plotone{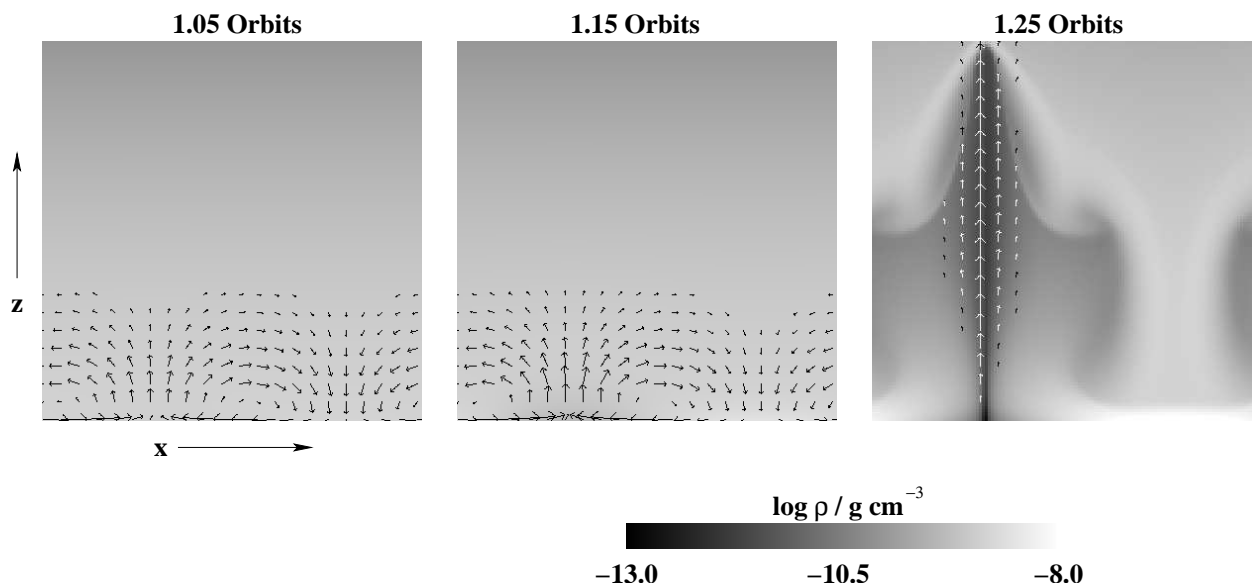}
\caption{Radiation hydrodynamical instability of the disk surface
layers in a 2-D calculation with a solid lower boundary of fixed
temperature.  The domain is centered a distance $H$ above the
midplane, has height and width $0.1 H$, and is divided into $128\times
128$ zones.  Random initial zone-to-zone density perturbations of one
part in $10^8$ are applied in the middle half of the domain height.
Results are shown at 1.05, 1.15 and 1.25~orbits (left to right).  Grey
shades indicate density on a shared scale logarithmic from $10^{-13}$
(black) to $10^{-8}$ g cm$^{-3}$ (white), and arrows show velocities.
The fastest speeds are $4\times 10^5$ cm s$^{-1}$ at left, $4\times
10^6$ at center, and $3\times 10^8$ at right.  Speeds in the center
panel are about equal to the initial isothermal gas sound speed at
domain center, $5.2\times 10^6$ cm s$^{-1}$.  The overturning pattern
is stationary until velocities exceed the gas sound speed, when a
chimney forms containing low-density gas driven upwards by a large
radiation flux.
\label{fig:overturn}}
\end{figure}

We search for growing radiation hydrodynamical modes of the surface
layers of the disk model shown in figure~\ref{fig:ic}, using numerical
calculations without magnetic fields.  The domain extends from $0.95$
to $1.05 H$ above the midplane, as described in
section~\ref{sec:icsurface}.  In the first series of calculations, the
domain width is equal to the height and symmetry is assumed along the
third dimension.  Initial density perturbations are applied in the
middle half of the domain height.  The perturbation in each grid zone
is random, with probability uniformly distributed over the interval
$-10^{-8}\leq\delta\rho/\rho\leq 10^{-8}$.  During the calculations an
exponentially-growing mode appears.  The horizontal variation can be
represented by a single Fourier component with wavelength equal to the
domain width while the vertical variation is the sum of several
Fourier components having growth rates identical within the time
fluctuations, indicating the mode is global.  The instability occurs
when low density at one location near the bottom boundary leads to a
larger radiation flux, which drives material up and away from the
boundary, further reducing the density.  In regions of higher density
to either side, gas falls, producing an overturning pattern shown in
figure~\ref{fig:overturn}.  The pattern is stationary and grows in
place.  It is localized near the bottom boundary and its amplitude
decreases sharply with height.  Density and radiation pressure are
approximately anticorrelated.  The largest positive density and
negative radiation pressure perturbations lie at the same horizontal
position, but the radiation pressure extremum is offset higher by~0.05
density scale height.  When the disturbance reaches non-linear
amplitudes, the rising gas forms an evacuated chimney in which the
flux is greater than the surroundings.  The horizontally-averaged flux
at the last time shown in figure~\ref{fig:overturn} is 4.6 times the
flux in the initial hydrostatic atmosphere.  Material is lost quickly
through the chimney, and the total mass in the domain decreases 4\%
during the 0.01~orbits before the last time shown.  The chimney
structure may prove to disrupt the layer in which it forms.  Radiation
diffusion is required for instability, as no growth is seen in an
otherwise identical calculation with the diffusion term
$-{\bf\nabla\cdot F}$ in equation~\ref{eqn:eoer} omitted.

\begin{figure}[ht!]
\epsscale{.6}
\plotone{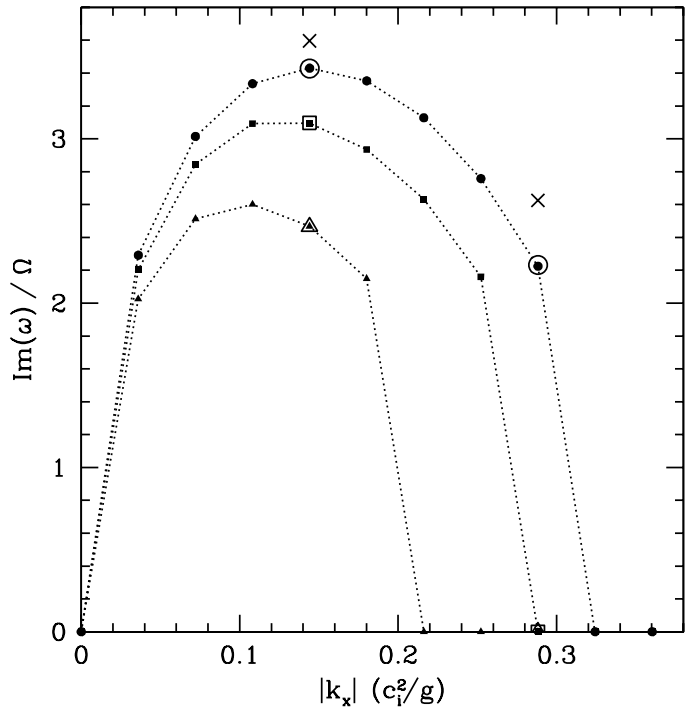}
\caption{Growth rate of the overturning instability versus horizontal
wavenumber in disk surface layer calculations.  The growth rates of
the $k_z=0$ Fourier components of the horizontal velocity are plotted
in units of the orbital frequency.  The unit of wavenumber is the
inverse gas pressure scale height at domain center.  Results from
calculations with domain size $0.4\times 0.1 H$ are shown by filled
symbols joined by dotted lines.  Results from calculations with domain
$0.1 H$ on a side are shown by larger open symbols.  The number of
zones in the domain height is 32 (triangles), 64 (squares), 128
(circles) or 256 (crosses).
\label{fig:shavivispectrum}}
\end{figure}

The variation of the instability with horizontal wavelength is checked
using calculations in a wider domain.  The width is increased fourfold
to $0.4 H$ and several modes with global vertical patterns grow
exponentially.  The dependence of growth rate on horizontal wavenumber
and numerical resolution is shown in figure~\ref{fig:shavivispectrum}.
The fastest modes have horizontal wavelengths roughly $0.1 H$.  Growth
rates increase with numerical resolution, so we have not achieved
numerical convergence.  The first null in horizontal velocity lies
2~grid zones above the bottom boundary in the calculations with 32
zones in the height, and 3, 6 and 12~zones above the boundary in the
calculations with 64, 128 and 256 zones in the height, respectively;
the pattern is poorly resolved on the coarser grids.  Like the type~I
instability, the overturning modes grow fastest at intermediate
horizontal wavenumbers, and are stable for $k_x=0$ and $k_x$ large.
In the linear analysis by \cite{sh01}, the fastest modes have
horizontal wavelengths roughly $2\pi$ times the density scale height
and growth rates comparable to the ratio of gas sound speed to density
scale height.  In the present calculations these correspond to
horizontal wavenumber $k_x c_i^2/g = 0.06$ and growth rate $1.3
\Omega$ at domain center.  The growth rate and wavenumber increase
with the Eddington ratio $\chi_F F/(cg)$ up to the largest value 0.9
examined by \cite{sh01}.  The Eddington ratio at domain center in our
calculations is 0.94 and the wavenumber and growth rate of the fastest
mode are greater than those estimated by Shaviv.

\begin{figure}
\epsscale{0.9}
\plotone{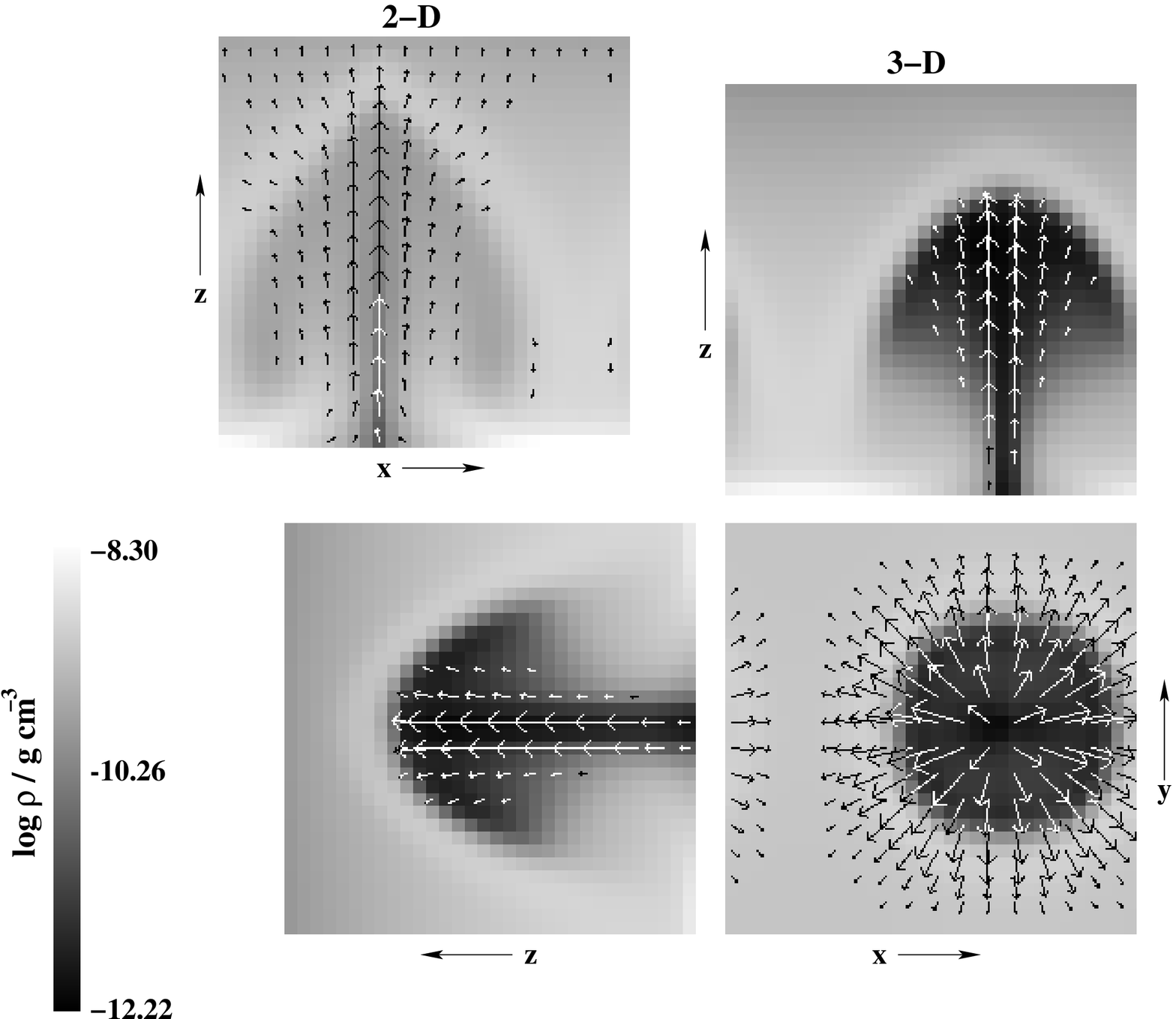}
\caption{Chimney structure resulting from the growth of the
overturning instability in 2-D (top left) and 3-D radiation
hydrodynamical calculations of the disk surface layers.  Grey shades
indicate densities, arrows velocities.  Results from the 2-D
calculation are shown at 1.54~orbits.  Slices through the 3-D
calculation at 1.58~orbits are perpendicular to the $y$-axis (top
right), $x$-axis (bottom left), and $z$-axis (bottom right).  The
vertical slices pass through the center of the chimney and the
horizontal slice is at domain center $z=H$.  There are 32 grid zones
along each direction.  The common logarithmic density scale spans
$6\times 10^{-13}$ (black) to $5\times 10^{-9}$ g cm$^{-3}$ (white).
The longest arrows show speeds of $8\times 10^7$ cm s$^{-1}$ in the
2-D calculation, $5\times 10^8$ in the vertical slices in the 3-D
calculation and $3\times 10^7$ in the horizontal slice.  For
comparison the initial radiation sound speed at domain center is
$9.5\times 10^7$ cm s$^{-1}$.
\label{fig:chimney}}
\end{figure}

Instability in three dimensions is tested using a single calculation
at the same location, with the volume of $(0.1 H)^3$ divided into
$32^3$ zones.  The fastest-growing mode is again an overturning
pattern that is global in the vertical direction, with wavelengths
along both horizontal axes equal to the domain size.  Its linear
growth rate $2.47 \Omega$ is the same within the amplitude of small
time variations as the fastest mode in the 2-D calculation of the same
domain width and resolution.  The gas motions soon after the
development of a chimney structure are shown in
figure~\ref{fig:chimney}.

The overturning instability appears to be of the first type described
by \cite{sh01}.  Both are local in the horizontal and global in the
vertical direction; require radiation diffusion; have density and
radiation pressure perturbations stationary and anti-correlated; and
grow fastest at about the gas sound crossing rate for wavelengths
comparable to the density scale height.  The instability is vigorous
in our disk surface layer calculations, growing faster than the
orbital frequency.  Disturbances reaching non-linear amplitudes
develop into narrow chimneys of low-density gas moving rapidly
upwards, separated by larger regions of denser down-welling material.
Radiation escapes readily through the chimneys.

No Shaviv type~I modes are observed in the convection calculation
shown in figure~\ref{fig:3up} where the spatial resolution matches
that in the surface-layer calculation with $32^2$ zones.  The type~I
modes may require a lower boundary of fixed temperature, and if this
is so they are unlikely to play a major role in cooling
radiation-dominated accretion disks.  The modes could be important in
situations where a radiation-supported atmosphere lies above a thermal
conductivity discontinuity, as in accreting, weakly-magnetized neutron
stars and in high-mass stars with internal composition boundaries.  No
modes resembling Shaviv's slower-growing type~II instability were
found in our calculations.  The absence may be due to differences
including our lower temperature and density, smaller gravity that
increases with height, and greater optical depth.

\section{LINEAR PHOTON BUBBLE INSTABILITY\label{sec:linear}}

Unlike convection and the overturning modes, photon bubbles require
magnetic fields.  The source of free energy for photon bubble
instability is the gradient in radiation pressure that supports the
disk against gravity in the vertical direction.  Instability occurs
when density disturbances lead to flux perturbations having a
component parallel to the magnetic field.  The greater flux in regions
of low density accelerates gas out of these regions along field lines,
leading to an increase in the density contrast with time as
illustrated in figure~\ref{fig:mechanism}.  A mathematical
justification for this picture is presented in the appendix.  When
magnetic fields are absent, compressional motions are purely
longitudinal in the short-wavelength limit.  Sound waves displace the
gas along the wavevector and perpendicular to the flux perturbations,
so there is no net acceleration over a wave period and no instability.

In this section we look at the exponential growth of small-amplitude
photon bubbles.  The growth rate varies with the wavelength, the
direction of propagation and the magnetic field strength and
orientation.  However the growth rate is independent of field strength
for magnetic pressures greater than the gas pressure.  Furthermore the
fastest modes have wavelengths shorter than $2\pi$ times the gas
pressure scale height but long enough to be optically-thick, and over
this range the growth rate varies slowly with wavelength.  The fastest
growth occurs when the wavefronts and the fields are close to the
vertical but tilted slightly with respect to one another.

\begin{figure}
\epsscale{0.6}
\plotone{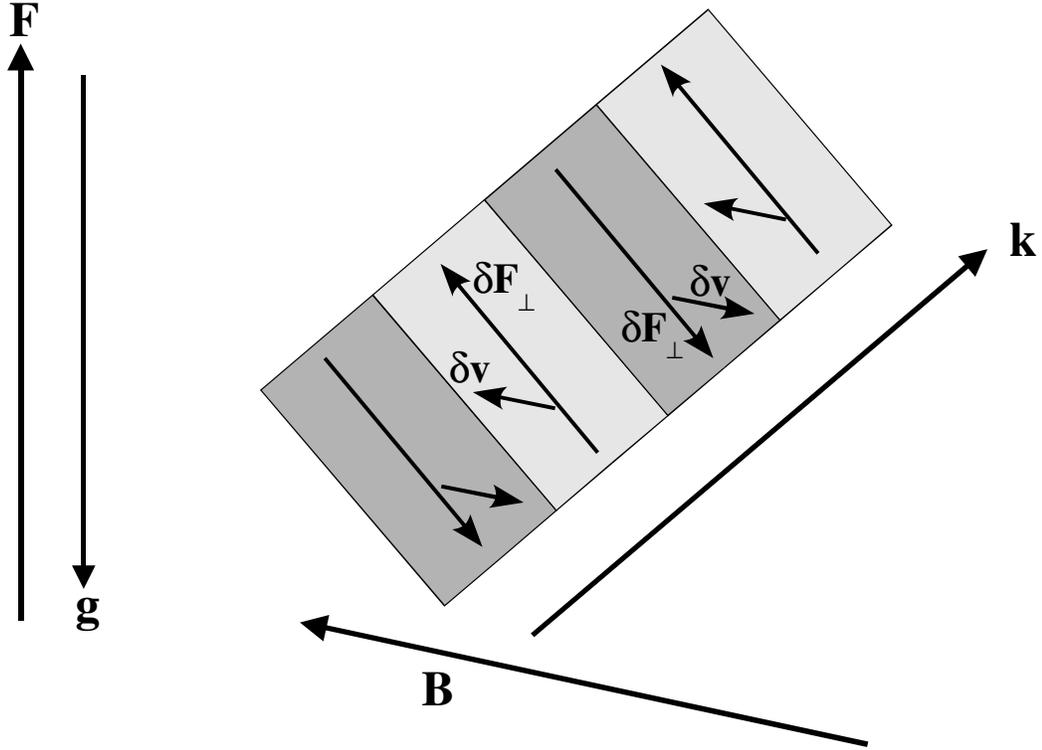}
\caption{Physics behind the radiative amplification of the slow
magneto\-sonic wave.  Fluctuations over two wavelengths are shown for
a plane wave propagating up and to the right with wave vector ${\bf
k}$.  The background upward radiative flux ${\bf F}$, downward
gravitational acceleration ${\bf g}$, and magnetic field ${\bf B}$ at
a randomly picked orientation are also shown.  Compressed and rarefied
regions are represented by dark and light grey, respectively, and are
oriented perpendicular to the wave vector.  The resulting diminished
and enhanced transparency of the medium produces a component of the
perturbed radiative flux, $\delta{\bf F}_\perp$, that is perpendicular
to the wave vector.  For the case where the magnetic energy density is
much larger than the gas pressure, the velocity perturbation
$\delta{\bf v}$ in the wave is aligned or anti-aligned with the
magnetic field, as shown.  In each region of the wave, there is a
nonzero, positive projection of $\delta{\bf F}_\perp$ onto $\delta{\bf
v}$.  This results in a radiative driving force that is always in
phase with the velocity, so the wave is amplified.
\label{fig:mechanism}}
\end{figure}

\subsection{Dependence on the Wavevector\label{sec:pbik}}

We compare the photon bubble growth against the predictions of
\cite{bs03}.  They carried out a WKB plane-wave analysis, treating the
radiation field in the Eddington approximation and found that photon
bubbles in the limit of short wavelengths are unstable slow
magneto\-sonic waves.  The growth rate increases with wavenumber for
fixed propagation direction, approaching an asymptote at wavenumbers
greater than the inverse gas pressure scale height $g/c_i^2$ if the
gas and radiation reach thermal equilibrium in much less than a wave
period.  The comparison is made using the same patch of the disk
surface layers as in section~\ref{sec:shavivi}.  In the first set of
2-D calculations, the magnetic field is initially uniform and
horizontal.  The field strength is chosen so the magnetic pressure is
10\% of the midplane radiation pressure, or 25\% of the radiation
pressure at the center of the surface layer patch.  Because the group
velocity is nearly parallel to the field, unstable waves arising near
domain center propagate many times across the width, taking more than
10~orbits to reach the upper or lower boundary.  Any possible effects
of the boundaries are further reduced by applying initial density
perturbations only in the middle half of the domain height.  As in
section~\ref{sec:shavivi}, the perturbations are random in each grid
zone, with $-10^{-8}\leq\delta\rho/\rho\leq 10^{-8}$.

\begin{figure}
\epsscale{0.75}
\plotone{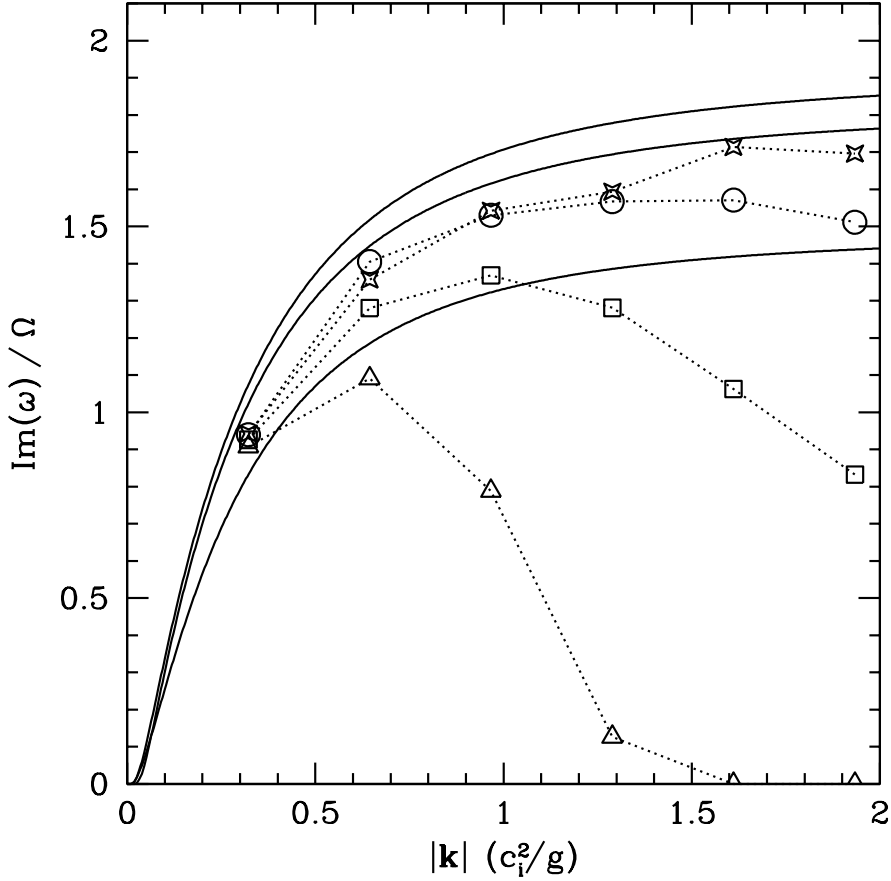}
\caption{Photon bubble growth rate versus wavenumber, in a small patch
of the disk surface layers with horizontal magnetic field having
pressure 10\% of the midplane radiation pressure.  The modes shown all
have wavevectors inclined $63^\circ$ from horizontal.  Solutions of
the \cite{bs03} linear dispersion relation are marked by solid curves,
results of numerical calculations by symbols.  The dispersion relation
was solved using conditions at the top, center, and bottom of the
patch (upper, middle, and lower solid curves, respectively).  The grid
resolutions in the numerical calculations are $32^2$ (triangles),
$64^2$ (squares), $128^2$ (circles) and $256^2$ zones (stars).  Growth
rates are plotted in units of the orbital frequency, wavenumbers in
units of the inverse gas pressure scale height at domain center.  The
linear analysis indicates growth rate is approximately independent of
wavenumber for $|{\bf k}| > g/c_i^2$, or wavelengths shorter than
$0.014 H$.  In the two highest-resolution calculations, the growth
rates of the four longest-wavelength modes agree well with the linear
analysis.
\label{fig:pbigrowth0}}
\end{figure}

During the evolution, mode amplitudes are measured every 0.01~orbits
using Fourier power spectra of the horizontal velocities.  The growth
rates are averaged in time over the period of exponential growth and
averaged together for pairs of modes with wavevectors mirror-symmetric
about the vertical.  The case of horizontal fields is special as modes
traveling to left and right grow at equal rates.  Growth rate is
plotted versus wavenumber in figure~\ref{fig:pbigrowth0}, for modes
with wavevectors $63^\circ$ from horizontal.  The longest-wavelength
of these modes fitting in the domain has one wavelength across the
width and two in the height and the first five overtones are also
shown.  The mean density scale height $0.037 H$ is intermediate
between the vertical wavelengths of the fundamental and first
overtone, so that the WKB assumption is violated for the fundamental
mode.  Exact numerical convergence of the surface-layer patch to a
single WKB solution is not expected because the variation of growth
rate with height depends on the wavenumber.  Nevertheless the measured
growth rates are in good agreement with the linear analysis.  Modes
that are well-resolved, having fifteen or more grid zones per
wavelength, grow at rates within 10\% of those predicted for domain
center.  The growth rate difference between the numerical solution and
the domain-center analytic solution for the second and third overtones
varies approximately quadratically with the grid spacing, as it should
with a numerical method of second-order accuracy.

\begin{figure}
\epsscale{0.7}
\plotone{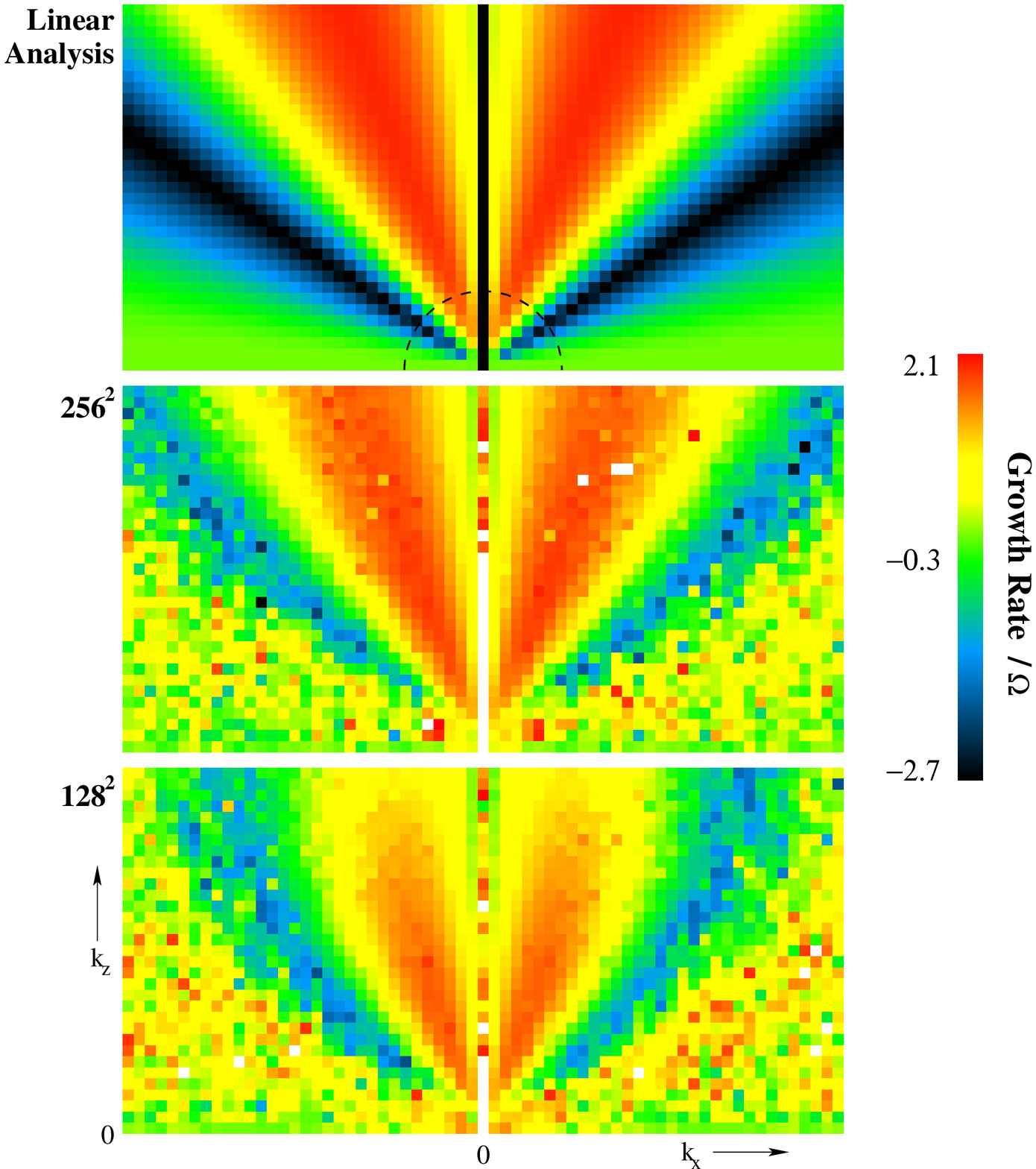}
\caption{Photon bubble growth rate (colors) versus horizontal and
vertical wavenumber in the disk surface layers.  The magnetic field is
horizontal, with pressure 10\% of the midplane radiation pressure.
Solutions of the \cite{bs03} linear dispersion relation for conditions
at domain center are shown at top.  Results of the highest-resolution
numerical calculations from figure~\ref{fig:pbigrowth0} are in the
lower panels.  The calculation with $256^2$ zones is shown at center,
and $128^2$ zones at bottom.  In each panel the origin $(k_x,
k_z)=(0,0)$ lies in the center bottom pixel.  Horizontal wavenumber
increases to the right, and vertical wavenumber increases upwards.
The top right corner corresponds to the mode with 32 wavelengths in
the domain width and 32 in the height, and wavevector inclined
$45^\circ$ clockwise from vertical.  The top left corner corresponds
to modes with the same wavelength, and wavevector $45^\circ$
counter-clockwise from vertical.  The shared color scale is linear
between $\leq -2.7\Omega$ (black) and $\geq 2.1\Omega$ (white).
Growth rate varies with propagation direction in about the same way
for all spatially-resolved wavenumbers greater than the inverse gas
pressure scale height $g/c_i^2$, marked in the top panel by a dashed
arc.
\label{fig:pbikxkz0}}
\end{figure}

The variation of growth rate with propagation direction is shown in
figure~\ref{fig:pbikxkz0}.  The most unstable modes at domain center
have wavevectors $68^\circ$ from horizontal.  Modes propagating
parallel to the radiative flux are stable because there is no flux
perturbation parallel to the wavefronts.  Modes propagating exactly
parallel or perpendicular to the magnetic field are stable because
displacements lie along the wavevector as in hydrodynamic acoustic
waves.  For other propagation directions, the growth rate increases
only slowly with wavenumber for $|{\bf k}| > g/c_i^2$, or wavelengths
less than one-seventh the domain height.  Lines of constant growth
rate in the top panel of figure~\ref{fig:pbikxkz0} are almost radial
except near $|{\bf k}| = 0$.

\subsection{Dependence on Timestep and Optical Depth\label{sec:pbitau}}

The effects of the numerical timestep on the linear growth rate are
checked using a version of the $128^2$ calculation with timesteps ten
times longer than the diffusion timestep.  The fundamental and first
overtone grow at rates similar to those shown in
figure~\ref{fig:pbigrowth0}, but the shorter-wavelength modes grow
more slowly than expected.  The shortest wavelength mode plotted in
figure~\ref{fig:pbigrowth0} grows at $1.5\Omega$ in the calculation
using the diffusion timestep, and $1.0\Omega$ in the calculation with
long timesteps.  These results are consistent with the expectation for
diffusion processes that the shortest wavelength accurately
represented is proportional to the square root of the timestep.

Photon bubble instability is absent in the limit of large optical
depth.  A version of the $32^2$ calculation is made with the diffusion
term $-{\bf\nabla\cdot F}$ omitted from equation~\ref{eqn:eoer}.
After 3~orbits, the largest speeds are $10^{-8}$ times the gas sound
speed, while by the same time in the version including diffusion,
speeds resulting from photon bubbles have grown greater than the gas
sound speed.

For sufficiently large wavenumbers, the optical depth per wavelength
is less than unity and disturbances in the radiation field are poorly
described by the Eddington approximation used in the linear analysis.
In the limit of low optical depth, there is no flux perturbation and
no photon bubble instability \citep{bs03}.  At domain center in the
surface layer calculations, the optical depth per wavelength is unity
at wavenumber $40 g/c_i^2$.  This wavenumber is unresolved even in our
highest-resolution $256^2$ numerical calculation, where grid zones at
domain center have optical depth 1.06.

Thermal equilibrium between gas and radiation holds throughout the
surface layer domain, as the time $t_{eqm}$ for the gas temperature to
change by emission and absorption of photons is much shorter even than
the oscillation period of the modes with unit optical depth per
wavelength.

\subsection{Three-Dimensional Modes}

\begin{figure}[ht!]
\epsscale{.6}
\plotone{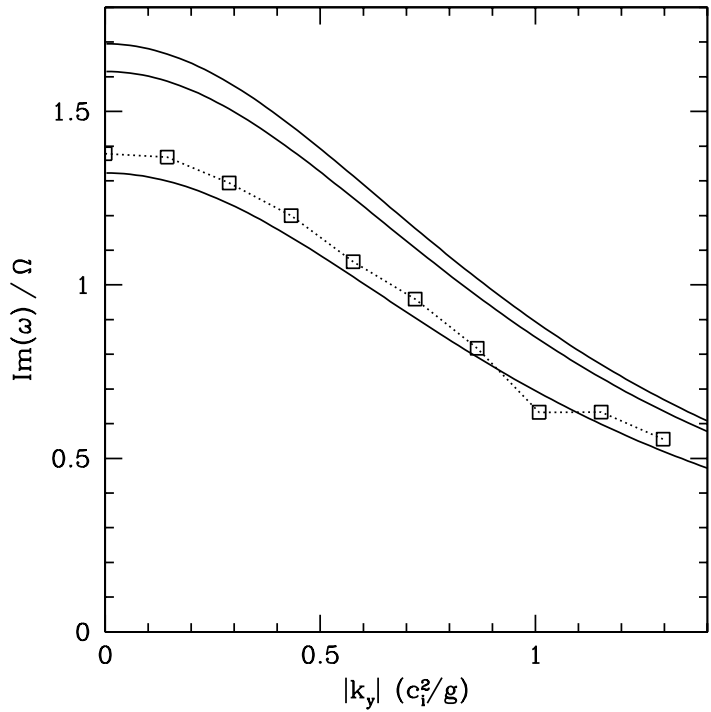}
\caption{Photon bubble growth rate versus wavenumber $k_y$ along the
third direction, in a patch of the disk surface layers with horizontal
$x$-magnetic field having pressure 10\% of the midplane radiation
pressure.  Squares indicate results from a numerical calculation of
$(0.1 H)^3$ of the surface layers, centered at $z=H$ and divided into
$64^3$ zones.  Solid curves mark solutions of the \cite{bs03}
dispersion relation for domain top, center and bottom.  The modes
plotted all have $x$-wavelength one-third the domain width and
$z$-wavelength one-sixth the domain height.  The symmetric mode with
$k_y=0$ also appears in figure~\ref{fig:pbigrowth0} as the second
overtone.  Growth rate decreases with wavenumber along the third
direction.
\label{fig:pbi3d1}}
\end{figure}

The fastest-growing linear photon bubble modes are symmetric
perpendicular to the plane containing the radiative flux and magnetic
field.  Modes with wavelength shorter than $0.1 H$ along the third
direction are expected to grow noticeably slower than the symmetric
modes in the surface layers of the model disk.  We carried out a 3-D
surface-layer calculation with a horizontal magnetic field having
pressure 10\% of the midplane radiation pressure.  The variation of
growth rate with wavenumber along the third direction is shown in
figure~\ref{fig:pbi3d1} and is consistent with the linear analysis.

\subsection{Dependence on the Magnetic Field}

\begin{figure}
\epsscale{0.75}
\plotone{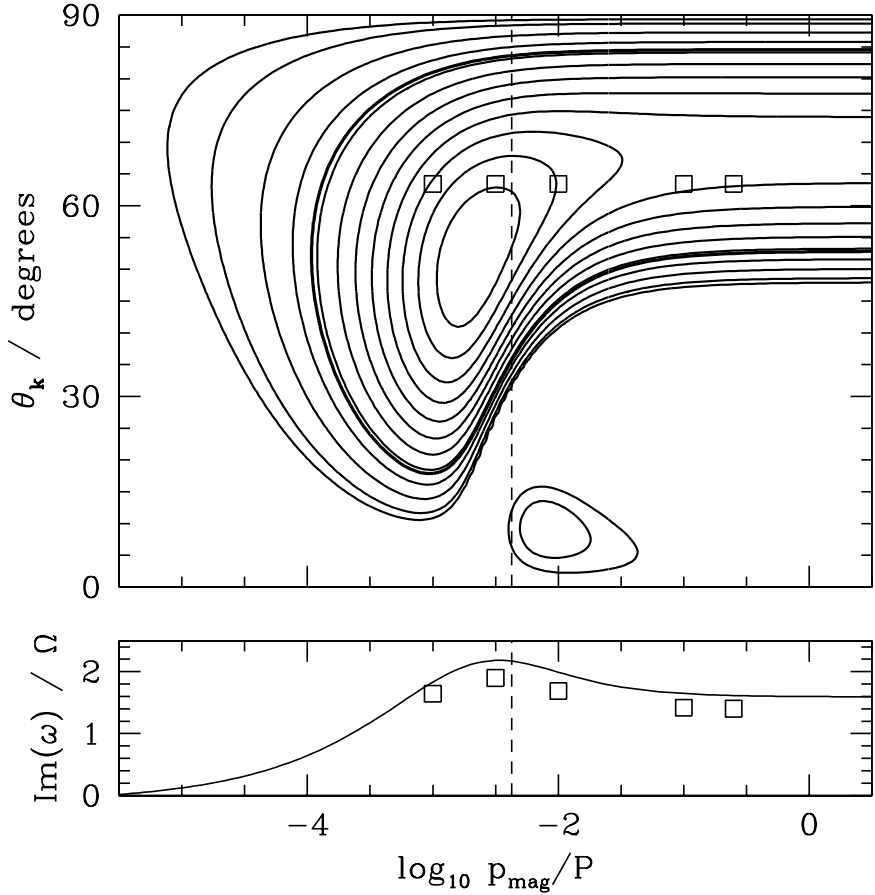}
\caption{Contours of photon bubble growth rate versus the magnetic
pressure and the angle between the wavevector and horizontal (top
panel).  The magnetic pressure is given on a logarithmic scale in
units of the radiation pressure at domain center $z=H$.  The
\cite{bs03} dispersion relation was solved for conditions at the same
location, assuming horizontal magnetic field and a wavenumber 97\% of
the inverse gas pressure scale height.  The chosen wavenumber is the
fastest growing in the $64^2$ calculation shown in
figure~\ref{fig:pbigrowth0}.  The growth rate at the lowest contour is
$0.1 \Omega$, and higher levels are 0.2, 0.4, $0.6 \Omega$ and so on.
A heavy contour marks growth rate $\frac{3}{4} \Omega$.  The fastest
growth rate $2.38 \Omega$ occurs at $(-2.65, 52^\circ)$.  Fast
magneto\-sonic waves are also weakly unstable, with fastest growth
rate $0.28 \Omega$ at $(-2.10, 9^\circ)$.  On horizontal magnetic
fields, the spectrum is mirror-symmetric about $\theta_{\bf
k}=90^\circ$ (figure~\ref{fig:pbikxkz0}).  Growth rate is plotted
against magnetic pressure in the bottom panel for the modes
propagating $63^\circ$ from horizontal.  Squares indicate results of
numerical calculations with $64^2$ grid zones.  The gas pressure is
shown by vertical dashed lines.  When magnetic pressure is greater
than gas pressure, the photon bubble growth rate is independent of
field strength.
\label{fig:pbipmag}}
\end{figure}

The spectrum of linear photon bubble modes is independent of magnetic
field strength if the magnetic pressure is greater than the gas
pressure, as shown in figure~\ref{fig:pbipmag}.  Growth at $z=H$ in
the model disk is faster than the MRI for some propagation direction
if the magnetic pressure is greater than one-thirtieth the gas
pressure.  The numerical results in figure~\ref{fig:pbipmag} are from
a set of 2-D surface-layer calculations in which the strength of the
initial horizontal magnetic field is varied.  In the cases with
magnetic pressure less than gas pressure, both the photon bubble and
overturn instabilities are present and gas motions lead to bending of
the fields.  Although the overturning modes are global, they are
localized near the bottom boundary, so photon bubble growth rates were
measured separately by using the horizontal velocities in the upper
half of the domain.  The overturning instability grows at $1.8 \Omega$
in the calculation with magnetic pressure 0.1\% of domain-center
radiation pressure, significantly slower than the rate $3.1 \Omega$ in
the calculation of the same resolution without magnetic fields shown
in figure~\ref{fig:shavivispectrum}.  The fields resist the
overturning motions.

\begin{figure}[t!]
\epsscale{0.6}
\plotone{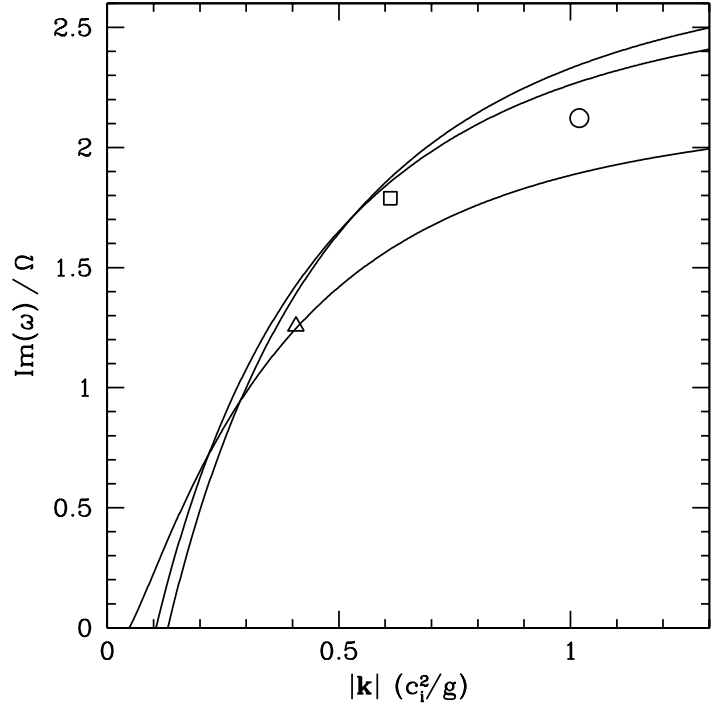}
\caption{Photon bubble growth rate versus wavenumber in a patch of the
disk surface layers with magnetic field inclined $12^\circ$ from
horizontal.  The triangle is from a run with $32^2$ zones, the square
from one with $64^2$ zones, and the circle from one with $128^2$
zones.  Only the fastest-growing mode in each run is plotted.  The
modes shown have wavefronts inclined $45^\circ$ from horizontal in the
same direction as the magnetic field.  The fastest-growing mode in the
$32^2$ calculation has horizontal and vertical wavelengths $0.05 H$,
or half the domain height.  Solid lines indicate solutions of the
linear dispersion relation for domain top, center and bottom.  Growth
rate increases toward the top of the domain for the highest
wavenumbers plotted.
\label{fig:pbigrowth12}}
\end{figure}

The growth rate of the photon bubble instability depends also on the
inclination of the magnetic field.  The effects of a small inclination
are examined using a set of 2-D surface-layer calculations, with a
magnetic pressure 10\% of the midplane radiation pressure as in
section~\ref{sec:pbik} but tilted $12^\circ$ from horizontal.  Growth
is fastest for wavefronts lying in the acute angle between the
magnetic field and the radiation flux.  The fastest modes have group
velocities lying close to magnetic field lines, with the vertical
component negative.  Photon bubbles originating near domain center
reach the bottom boundary after crossing the width of the box about
2.4~times, in 0.8~orbits.  The reflecting lower boundary leads to some
exchange of energy between modes, so that only the growth of the
fastest mode is reliably measured.  The linear growth rates of the
fastest modes in calculations of different resolutions are plotted
against wavenumber in figure~\ref{fig:pbigrowth12}.  The differences
from the linear analysis are less than the variation in growth rate
over the domain height.

\subsection{Fastest-Growing Modes\label{sec:pbifast}}

\begin{figure}
\epsscale{0.85}
\plotone{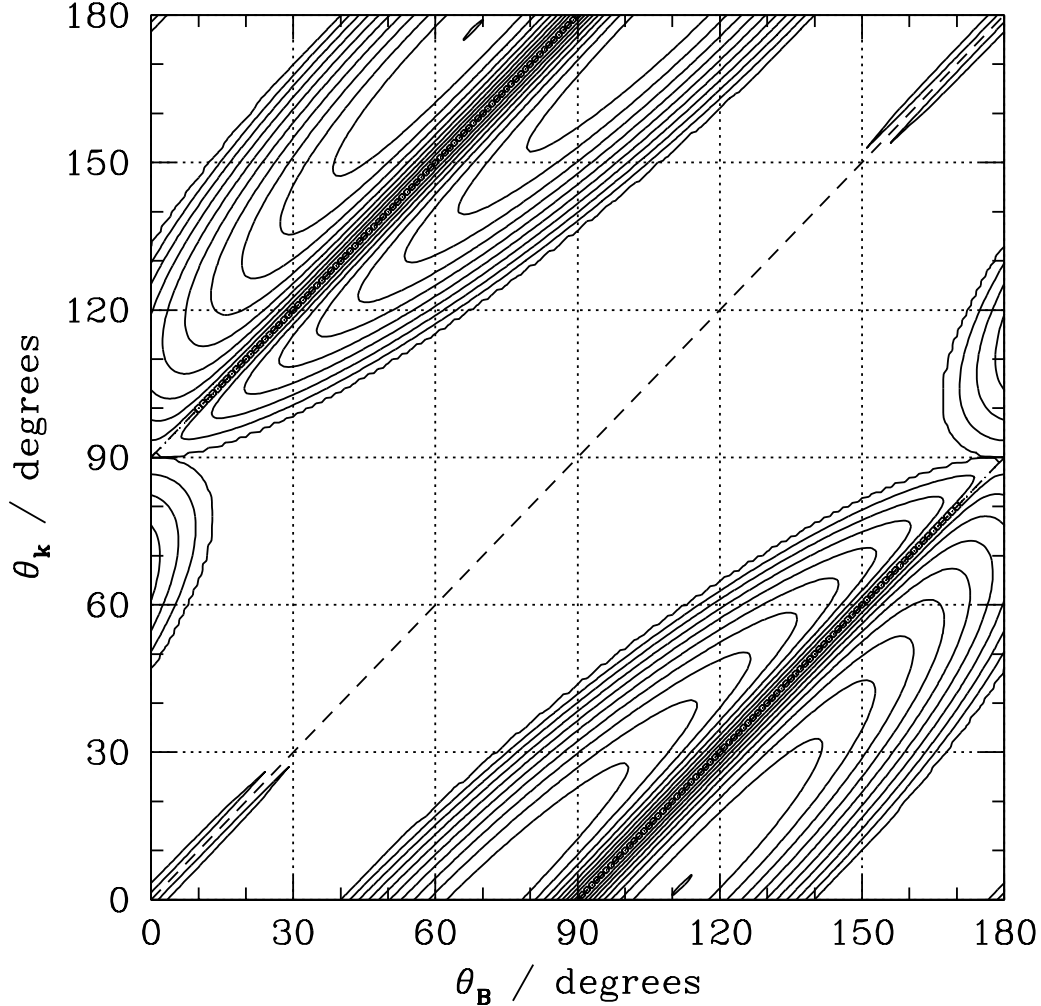}
\caption{Contours of the growth rate as a function of the inclinations
of the magnetic field and wavevector.  The \cite{bs03} dispersion
relation was solved at domain center in the surface layer
calculations, in the limit of a strong magnetic field and for
wavenumber $g/c_i^2$.  The lowest contour is at $0.015\Omega$ and the
remainder are at integer and half-integer multiples of the orbital
frequency.  The fastest growth rate $5\Omega$ occurs for magnetic
fields $68^\circ$ and wavevectors $177^\circ$ from horizontal.  Dashed
diagonal lines indicate wavevectors parallel and perpendicular to the
field.  Photon bubbles are slow magneto\-sonic modes and grow fastest
for wavevectors nearly perpendicular to the field.  The unstable waves
propagate at or above the horizontal.  Fast magneto\-sonic modes
propagating almost parallel to the field are also unstable but grow
slowly, as shown by the contours nearest the origin.  Growth rates are
identical for field angles $180^\circ$ apart.
\label{fig:pbiangles}}
\end{figure}

Photon bubble growth rates depend on the orientations of the
wavevector and magnetic field.  The angles of fastest growth are found
in this section.  The results are independent of the field strength,
as the magnetic pressure is fixed at 10\% of the midplane radiation
pressure and is much greater than the gas pressure.  Based on the
linear analysis and results of sections~\ref{sec:pbik}
and~\ref{sec:pbitau}, growth is expected to be fastest for wavenumbers
between $g/c_i^2$ and $2\pi\chi_F\rho = 40 g/c_i^2$.  The angle
dependence of the growth rates at the low end of this range is shown
in figure~\ref{fig:pbiangles}.  The growth rates of photon bubbles
with wavenumber equal to the inverse gas pressure scale height can be
seen to increase with the inclination of the magnetic field.  The
fastest mode on horizontal fields grows at $1.73 \Omega$, on fields
inclined $3^\circ$ at $2.07 \Omega$ and on fields inclined $12^\circ$
at $2.95 \Omega$.  The fastest modes of all have wavevectors almost
horizontal, magnetic field inclined $22^\circ$ from vertical in the
opposing sense and growth rate $5.00 \Omega$.  The patterns are
similar for larger wavenumbers of 10 and 40~times the inverse gas
pressure scale height except that the fastest modes have wavevectors
more nearly horizontal and magnetic fields more nearly vertical so
that the perturbations in flux and velocity are close to parallel and
driving is maximized.  The growth rates are 8.9 and 9.5 times the
orbital frequency, respectively.  The maximum growth rate at the
largest wavenumber is 9\% less than the asymptotic rate $g/(2 c_i) =
10.4 \Omega$ obtained from \cite{bs03} equation~93 with radiation and
magnetic pressures much greater than gas pressure.

Accurate measurements of growth rate on inclined fields are difficult
in surface-layer calculations owing to the effects of the lower
boundary.  We measure growth rates instead using a higher-resolution
version of the radiation-MHD calculation shown in
figure~\ref{fig:3up}, spanning the disk thickness.  The grid
resolution is doubled to $256\times 1472$ zones so that the grid
spacing is the same as in the $64^2$ surface-layer runs of
figures~\ref{fig:pbigrowth0} and~\ref{fig:pbigrowth12}.  As in
figure~\ref{fig:3up}, the timesteps chosen are 100~times the diffusion
step $\Delta t_D$ set by the minimum density, found outside the
photosphere.  More precise measurement of the linear growth rates is
made possible by smaller initial random density perturbations of one
part per million.  The mode amplitudes are measured from the Fourier
transforms of the horizontal velocity in the square region between
0.65 and $1.05 H$ above the midplane.  The fastest mode showing
sustained exponential growth has 31 wavelengths in the width and 8 in
the height of the selected region, corresponding to a wavevector
$14^\circ$ from horizontal, a wavenumber 1.16 times the initial
inverse gas pressure scale height at $z=H$ and a wavelength of 8~grid
zones.  Its growth rate is $5.00 \Omega$.  The local dispersion
relation near the top of the region at $z=H$ indicates this mode is
expected to grow at $5.74 \Omega$, fastest among all those with the
same and smaller wavenumber.  The dispersion relation near the bottom
of the region at $z=0.7 H$ indicates the same mode is expected to grow
at $4.07 \Omega$.  The growth rate measured in the numerical
calculation lies near the middle of the range expected over the region
based on the linear analysis.

\section{SHOCK TRAINS\label{sec:shocks}}

\cite{be01} showed that a radiation-supported atmosphere with a strong
magnetic field can sustain a train of propagating shocks.  A 1-D
analytic solution was found by neglecting the change in background
quantities from one shock to the next so that the flow is periodic and
assuming rapid photon diffusion so the gas is isothermal.  In the
solution, Lagrangean fluid elements move back and forth along inclined
magnetic field lines.  Where the density is low, the gas transmits a
large radiative flux and is driven up field lines by the radiation
force.  On striking high-density material it shocks, is compressed and
slides back down the field under gravity.  The solution relates the
inclination and spacing of the fronts, the density jump across the
shocks and the overall flux of radiation through the atmosphere.
Given any two of these four quantities, the other two can be
calculated.

The shock train, like the photon bubble instability, is a propagating
disturbance that requires a magnetic field and is driven by radiative
flux changes associated with density perturbations.  In this section
we show that photon bubbles reaching non-linear amplitudes become
shock trains.  To spatially resolve the fast-growing linear modes with
wavenumbers near the inverse gas pressure scale height, we use
calculations of the disk surface layers.  A fiducial calculation is
described in section~\ref{sec:shockfiducial} and compared with the
non-linear analytic solution in section~\ref{sec:shocksolution}.  The
stability of the shocks is discussed in
section~\ref{sec:shockstability} and the dependence on magnetic field
strength and orientation in section~\ref{sec:shockb}.  Limits on
growth near the disk photosphere are examined in
section~\ref{sec:shocktau}.  The wavelengths of disturbances reaching
order unity generally increase with time.  Calculations extending
through the whole thickness of the disk have lower resolution, but
allow the shocks to become more widely-spaced and better show the
effects of the vertical gradients in background quantities.  They are
described in section~\ref{sec:shockfull}.

\subsection{Fiducial Calculation\label{sec:shockfiducial}}

\begin{figure}[t!]
\epsscale{0.8}
\plotone{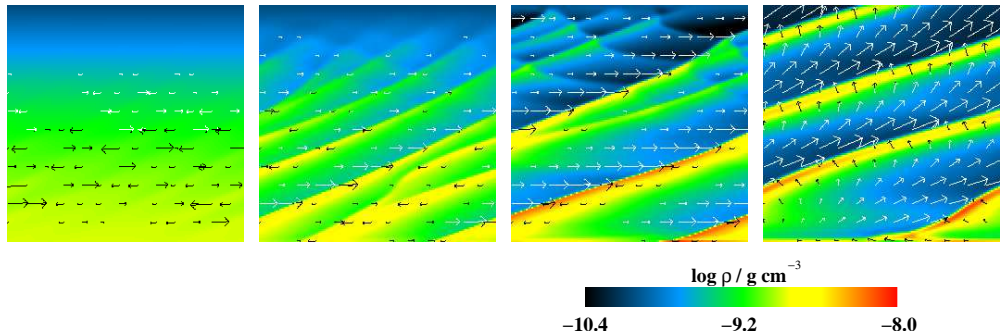}
\caption{Time sequence showing shock train development in the fiducial
calculation of the disk surface layers.  The magnetic field is
initially $3^\circ$ from horizontal and has pressure 25\% of the
radiation pressure at domain center.  Results are shown at 0.7, 1.1,
1.5 and 1.77~orbits (left to right) with densities marked by colors on
a common logarithmic scale and velocities by arrows.  The longest
arrows correspond to speeds of $4\times 10^5$, $1\times 10^7$,
$2\times 10^7$ and $6\times 10^7$ cm s$^{-1}$ (left to right).  The
first panel falls at the end of the linear phase, the second in the
shock merger phase, the third near the final shock merger, and the
fourth at the start of strong outflow through the top boundary.
\label{fig:shockfiducial}}
\end{figure}

A fiducial calculation of the growth of shocks in the disk surface
layers is made using conditions identical to those in
section~\ref{sec:pbik} except that the field is tilted $3^\circ$ from
horizontal and the random initial density perturbations have amplitude
0.1\%.  The domain is divided into $128\times 128$ zones.  The
calculation passes through three main phases: exponential growth,
merging shocks and a steady shock pattern
(figure~\ref{fig:shockfiducial}).  During an initial transient lasting
0.1~orbits, the density perturbations lead to disturbances in the flux
and to gas motions.  Photon bubble modes then become established and
grow exponentially.  Modes with wavefronts tilted to the same side as
the magnetic field grow fastest.  The most unstable has 3 wavelengths
in the domain width and 7 in the height and grows at $1.89 \Omega$.
Velocities exceed 10\% of the isothermal gas sound speed at
0.72~orbits and the wave pattern steepens into a train of traveling
shocks, with the same orientation as the wavefronts of the fastest
linear mode.  Growth slows at 0.9~orbits, when densities range from
half to twice their initial values.  During the second phase, shocks
merge because gas accelerated through the low-density regions reaches
lower speeds if the fronts are closer together.  A trailing shock
propagates faster into the slow-moving upstream gas, eventually
overtaking the shock ahead.  The merged fronts are more and more
widely spaced and the density contrast and mean radiation flux
increase.  Mergers continue until at 1.6~orbits there is just one
front in the domain width and two in the height.  The horizontal
spacing of fronts is then constant during the third phase.
Propagation is fastest near the bottom edge where the local flux is
greatest due to the fixed temperature of the boundary.  The fronts are
sheared and become more nearly horizontal over time.  The mean flux
through the top boundary and the gas speeds increase until at
1.74~orbits ram pressure exceeds magnetic pressure at some locations,
the fields buckle, the pattern is disrupted and material is lost
through the top boundary.

\subsection{Comparison with Analytic Solution\label{sec:shocksolution}}

We compare the results of the fiducial calculation against an improved
version of the \cite{be01} shock train solution.  Because the flux is
proportional to the radiation energy gradient divided by the density,
the curl of $\rho{\bf F}$ ought to be zero.  This is ensured by
choosing a form for the flux
\begin{equation}
{\bf F} = {\bf F}_0 - \frac{c g}{\chi_F}
\left(\frac{\rho-\rho_0}{\rho}\right) \sin\theta_f
\left({\bf\hat f} + \frac{4 E v_0 \chi_F}{3cg} {\bf\hat k}\right),
\end{equation}
replacing equation~16 of \cite{be01}.  The first term ${\bf F}_0 =
{\bf\hat z}cg/\chi_F$ is the flux through the hydrostatic atmosphere.
The disturbances are separated into components along the unit vector
$\bf\hat f$ parallel to the shock fronts and along the perpendicular
direction $\bf\hat k$.  We write $\theta_f$ for the angle between the
shock fronts and the horizontal, $\rho_0$ for the mean density and
$v_0$ for the horizontal component of the shock propagation speed.
The form of the flux used by \cite{be01} when written in this notation
differs only by an extra factor $\sin^{-2}\theta_f$ in the term
proportional to $\bf\hat f$.  The solution is obtained by the same
steps used in \cite{be01} and relates the same four parameters, the
inclination and spacing of the shock fronts, the size of the density
jump and the enhancement in the flux.  The horizontal distance between
shocks in units of the gas pressure scale height is
\begin{equation}\label{eqn:lambda}
\lambda_x = \left(\cot(\theta_f-\theta_B) -
{{\left|\sin(\theta_f-\theta_B)\right|}\over{\sin\theta_f}}
{4Ec_i\chi_F\over 3cg} \right)^{-1}
{{\eta_+ - \eta_+^{-1}+2\ln\eta_+}\over{\sin^2\theta_f}}
\end{equation}
and the enhancement in the vertical component of the flux in the limit
of strong magnetic fields becomes
\begin{equation}\label{eqn:ell}
\frac{F_z}{F_0} = 1 + \left(\sin^2\theta_f +
\left|\sin(\theta_f-\theta_B)\right| \cos\theta_f
\frac{4Ec_i\chi_F}{3cg}\right)
\frac{\eta_+^2-\eta_+^{-2}-4\ln\eta_+}
{2(\eta_+ - \eta_+^{-1} + 2\ln\eta_+)},
\end{equation}
where $\eta_+$ is the ratio of the maximum density to the mean and
$\theta_B$ is the angle between the magnetic field and the horizontal.
The ratio $\eta_-$ of the minimum density to the mean is expected to
equal $\eta_+^{-1}$.  The quantities we compare between the analytic
and numerical solutions are (1) the density structure, (2) the density
contrast -- shock spacing relationship and (3) the flux enhancement.

\begin{figure}
\epsscale{0.75}
\plotone{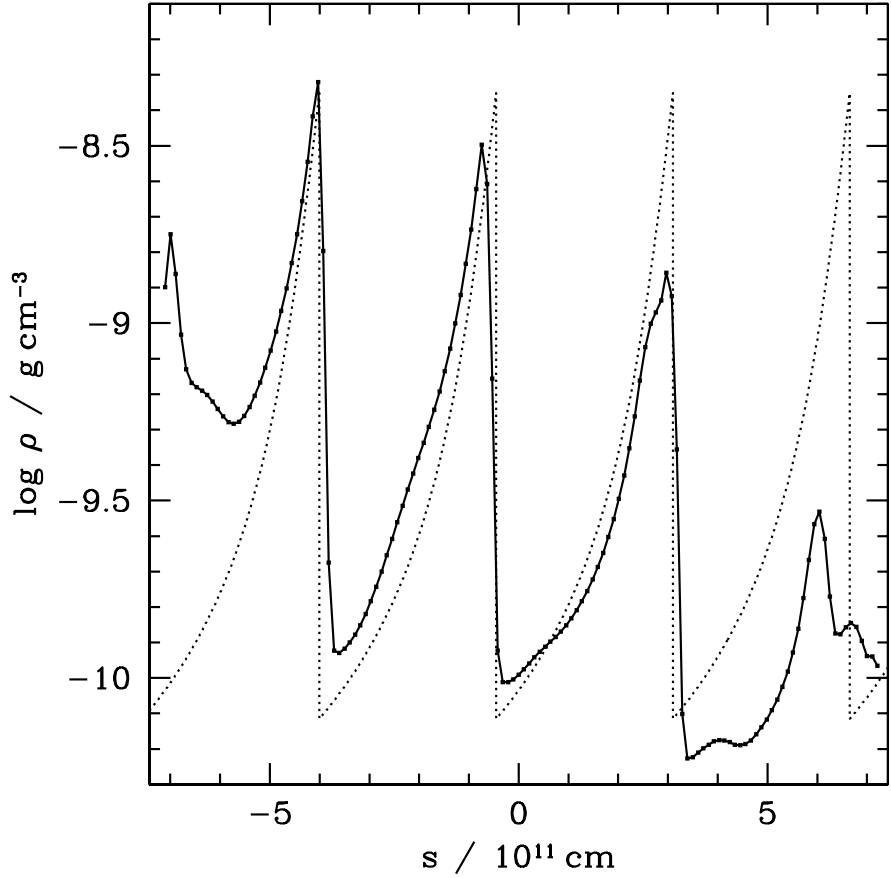}
\caption{Density along a ray through domain center perpendicular to
the shock fronts at 1.7~orbits in the fiducial calculation of a patch
of the disk surface layers.  The horizontal coordinate $s$ is the
distance along the ray.  The lower boundary $z=0.95 H$ lies at left
and the upper boundary $z=1.05 H$ at right.  The numerical results are
shown by points and a solid line, the non-linear analytic solution
with the same shock spacing and inclination by a dotted line.
\label{fig:shockcut}}
\end{figure}

The density along a ray perpendicular to the shock fronts at domain
center in the fiducial calculation is shown with the analytic solution
in figure~\ref{fig:shockcut}.  For the analytic solution, the
inclination and spacing of the fronts are set equal to the values
measured in the numerical calculation and the amplitude of the density
variation is found by inverting equation~\ref{eqn:lambda}.  There is
general agreement between the two solutions in the amplitude and shape
of the resulting pattern, but several differences are apparent.  The
analytic solution is periodic by assumption, while the background
density and shock strength in the numerical calculation vary with
height and are greater below domain center.  The pattern is reduced in
strength near the top and bottom boundaries.  Also, the shocks are
spread over several grid zones in the fiducial calculation for
numerical stability, while in the analytic solution the shocks are
arbitrarily thin.  Similar profiles are found on other rays
perpendicular to the fronts.  The agreement in the fiducial
calculation is better than in a version with a coarser grid of $64^2$
zones, indicating that some of the departure from the analytic
solution is due to the limited spatial resolution.  At earlier times
during the shock merger phase in the fiducial run, a range of
separations is present, violating the analytic assumption of periodic
structure.

\begin{figure}
\epsscale{0.75}
\plotone{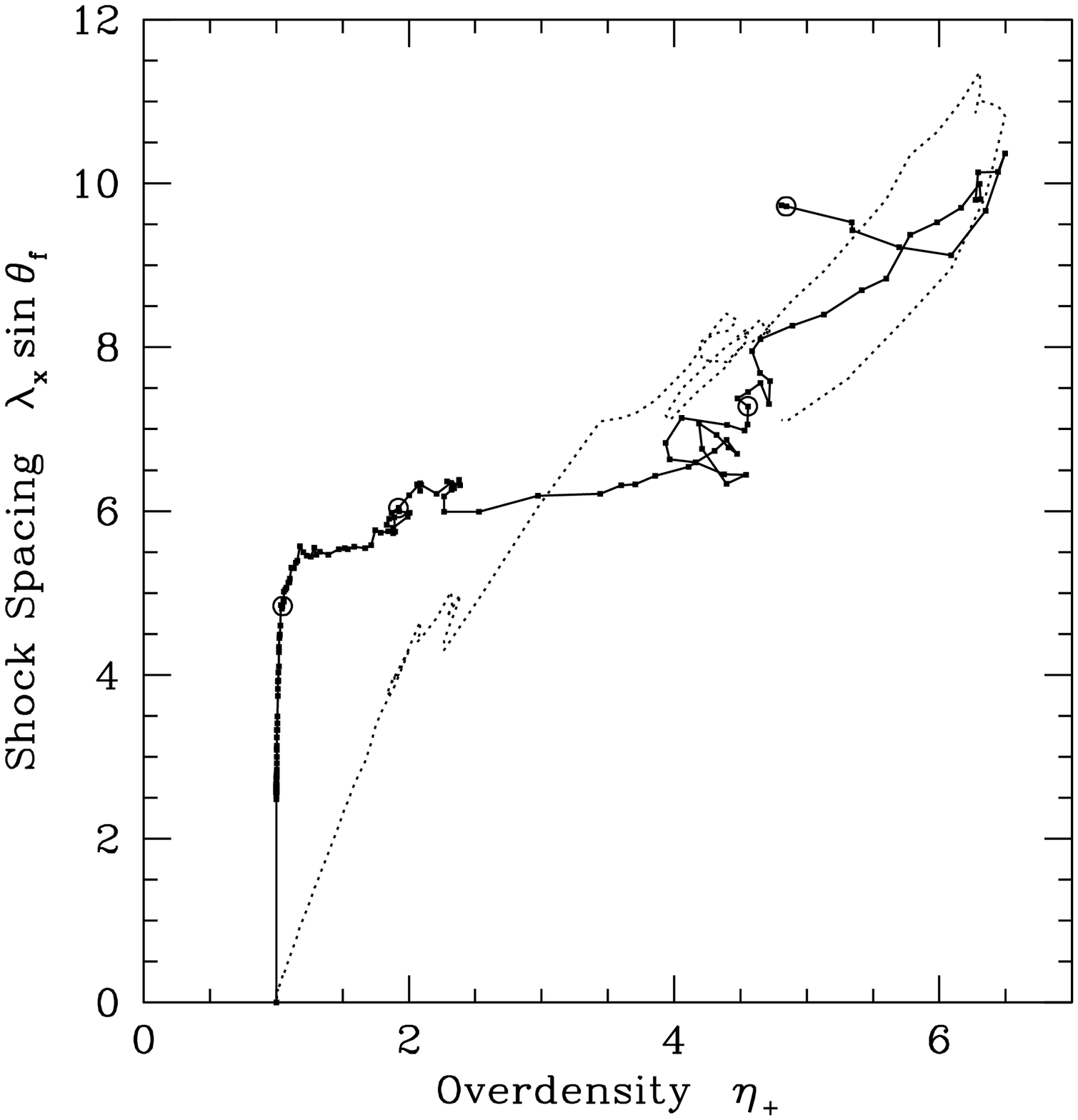}
\caption{Shock separation versus overdensity in the disk surface
layers.  The perpendicular distance between shocks is plotted in units
of the gas pressure scale height at domain center.  Results from the
fiducial calculation are shown by points joined by a solid line.  Time
generally increases toward the top right and the points are separated
by 0.01~orbits.  Open circles mark the four times shown in
figure~\ref{fig:shockfiducial}: 0.7, 1.1, 1.5 and 1.77~orbits.
Corresponding analytic estimates calculated with the overdensity and
shock orientation in the numerical calculation using
equation~\ref{eqn:lambda} are shown by a dotted line.
\label{fig:shockspacing}}
\end{figure}

The distance between shocks in the fiducial calculation increases with
the overdensity as shown in figure~\ref{fig:shockspacing}.  The
domain-averaged shock spacing and inclination are measured by locating
local density maxima along rows and columns.  Maxima found next to
converging flows are counted as shocks.  Also shown is an analytic
estimate computed from equation~\ref{eqn:lambda} using the
overdensities and shock inclinations in the numerical calculation, the
initial magnetic field angle and the background quantities at domain
center.  The relationship between overdensity and shock separation in
the numerical calculation agrees well with the analytic expectation
except near the start and end of the calculation.  Near the start, the
front spacing is overestimated because the small-amplitude waves have
density maxima offset by a quarter-wavelength from the locations of
fastest compression and are missed from the count.  Near the end, the
analytic assumptions of straight magnetic field lines and
time-averaged hydrostatic equilibrium no longer hold.

\begin{figure}
\epsscale{0.75}
\plotone{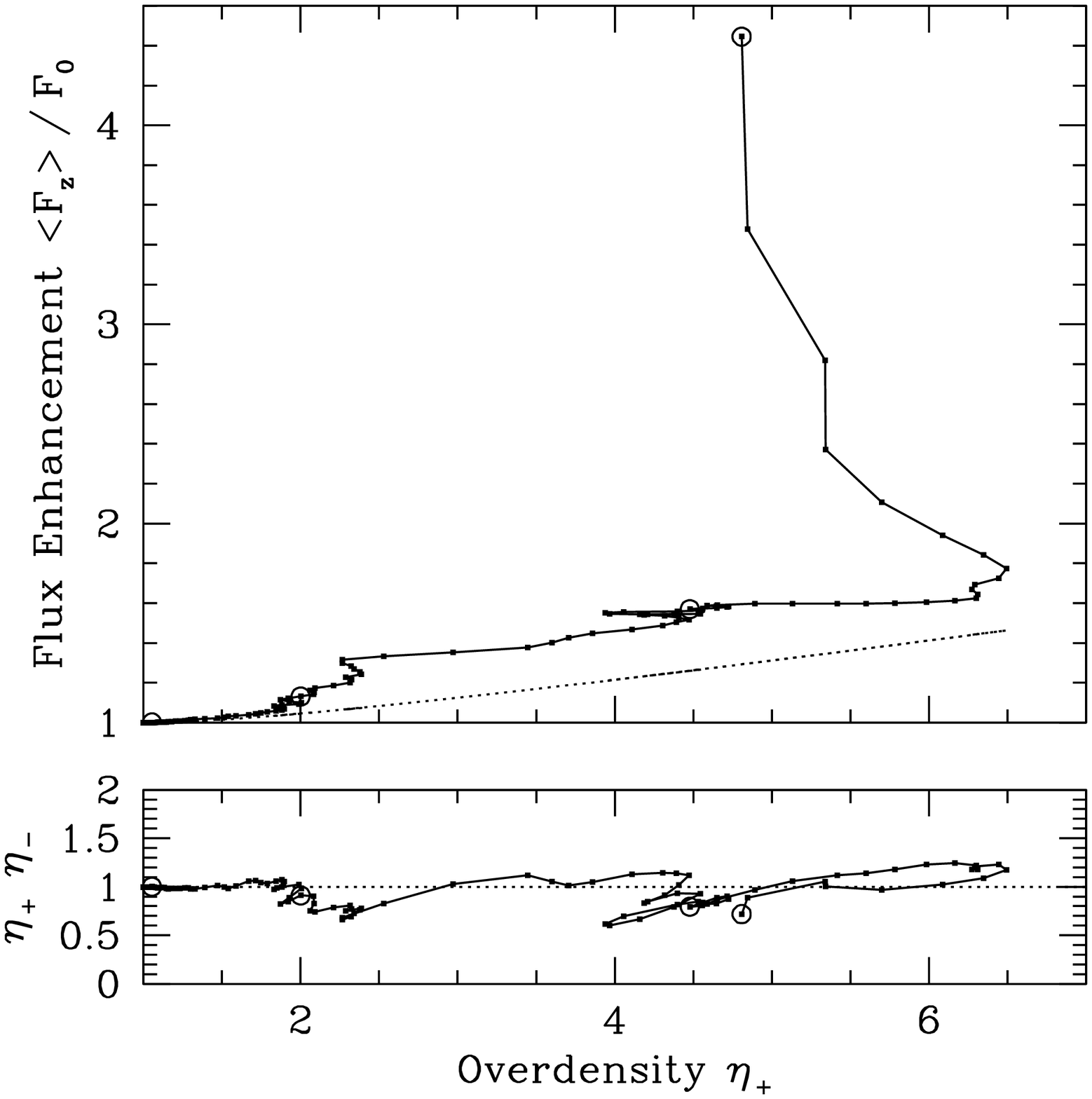}
\caption{Enhancement of the mean radiation flux over the hydrostatic
value, as a function of the overdensity.  Measurements from the
fiducial numerical calculation are shown by points joined by a solid
line.  The relationship expected from the shock train analysis
(equation~\ref{eqn:ell}) is shown by a dotted line.  Time increases
toward the right and top and adjacent points are separated by
0.01~orbits.  Open circles mark the four times 0.7, 1.1, 1.5 and
1.77~orbits shown in figure~\ref{fig:shockfiducial}.  The analytic and
numerical results agree closely until 1.1~orbits when the product of
under- and overdensities, plotted in the bottom panel, first departs
from the value unity expected in the analytic theory.  The fluxes in
the two solutions then grow at similar rates until 1.7~orbits despite
increasing violation of the analytic assumption that the shock spacing
is much less than the density scale height.  After 1.7~orbits,
magnetic fields buckle in the numerical calculation and fluxes are
much greater than in the analytic solution.
\label{fig:shockflux}}
\end{figure}

The radiation flux in the fiducial calculation is shown as a function
of the overdensity in figure~\ref{fig:shockflux}.  The vertical
component of the flux in the frame co-moving with the gas is averaged
along a horizontal line passing through domain center, and plotted in
units of the flux in the initial hydrostatic atmosphere.  The
overdensity is the maximum ratio of the density to the initial density
on the same horizontal line.  Each plotted point marks a measurement
from one snapshot in the fiducial calculation, while a dotted curve
shows the analytic estimate from equation~\ref{eqn:ell}.  The two
solutions agree well until the shock spacing approaches the density
scale height, violating the analytic assumption that the background
quantities vary little over a wavelength.  The flux in the numerical
calculation is then greater, but increases with overdensity at about
the same rate as in the analytic solution.  During the final
0.07~orbits, fluxes become much greater than in the analytic solution
as the magnetic fields buckle and gas is ejected through the top
boundary.

\subsection{Stability in Three Dimensions\label{sec:shockstability}}

\begin{figure}
\epsscale{0.75}
\plotone{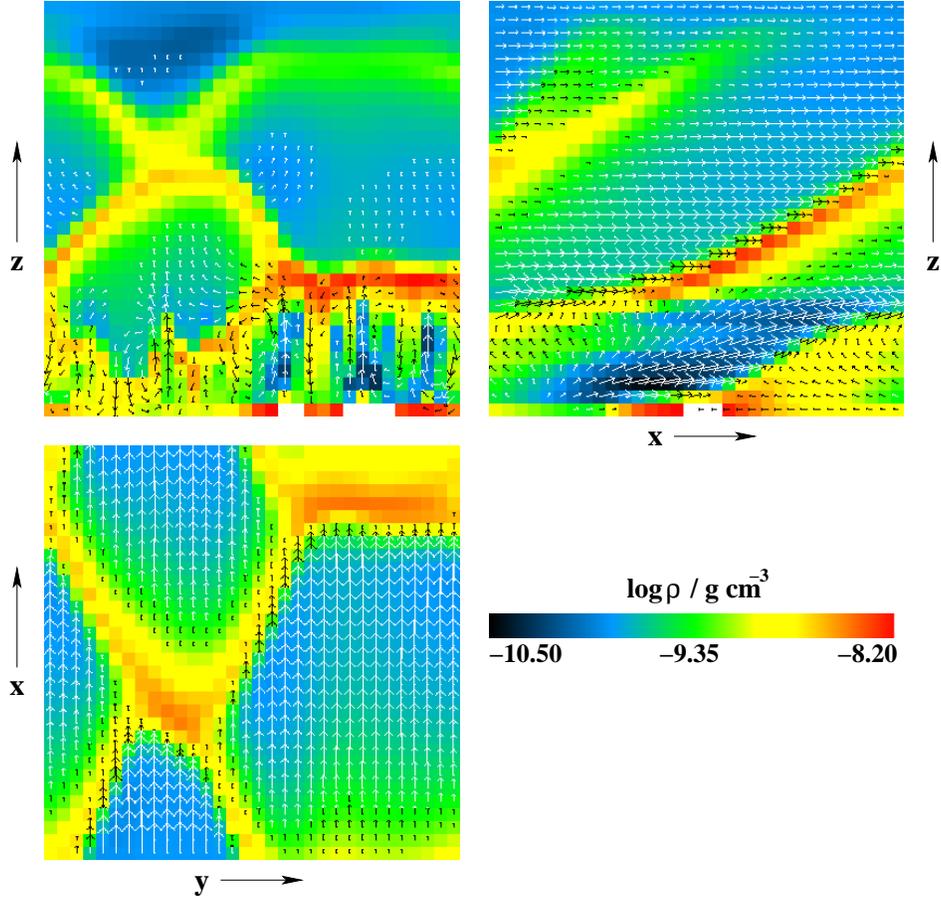}
\caption{Shock pattern and secondary instability in a 3-D calculation
of the disk surface layers.  The shocks form the walls of an inclined
honeycomb pattern opening toward domain top.  Shown at 1.65~orbits are
cross-sections perpendicular to the $x$- (top left), $y$- (top right)
and $z$-axes (bottom left).  Colors indicate densities, arrows
velocities.  The longest arrows correspond to speeds of $6\times
10^6$, $3\times 10^7$ and $3\times 10^7$ cm s$^{-1}$, respectively.
The top right panel cuts the $y$-axis 80\% of the way across the
domain, at a position where the secondary instability leads to large
upward speeds near the bottom boundary.  The top half of the panel
resembles the 2-D fiducial results in figure~\ref{fig:shockfiducial}.
The cross-sections in the other two panels pass through domain center.
The pattern of thin slabs of low-density rising gas due to the
secondary instability can be seen near the domain floor in the top
left panel.  As in the fiducial calculation, the magnetic field lines
are initially inclined $3^\circ$ from the positive $x$-axis toward the
$z$-axis.
\label{fig:shock3d}}
\end{figure}

Shocks grow in the fiducial calculation until the pattern is disrupted
by buckling of the magnetic field.  However in three dimensions, the
field provides stiffness only along one horizontal axis, and
variations along the other horizontal direction could destroy the
shock pattern at an earlier stage.  A three-dimensional version of the
fiducial calculation is made by extending the domain along the third
or $y$-axis to make a cube of volume $(0.1 H)^3$ divided into $32^3$
zones.  Other parameters are identical to the fiducial calculation,
with each magnetic field line lying on an $x$--$z$ plane and inclined
$3^\circ$ from horizontal.  The results on individual cross-sections
perpendicular to the third direction are initially quite similar to
the fiducial run, with the linear instability leading to a train of
parallel inclined propagating shocks that merge, until at 1.5~orbits
the horizontal shock separation is equal to the domain $x$-extent.
Stucture along the third axis is at large scales, with the greatest
Fourier power in the modes with wavelength equal to the domain
$y$-size.  The development is changed by a secondary instability that
becomes apparent after 1.5~orbits, when an overturning pattern appears
near the bottom boundary (figure~\ref{fig:shock3d}).  The overturning
motions lie in the $y$--$z$ plane and are approximately symmetric
along the $x$-direction parallel to the magnetic field.  Some field
lines move up, leaving behind lower gas densities near the bottom
boundary, while field lines at nearby $y$-positions move down, locally
increasing the density.  The wavelength along the $y$-direction is
about one-sixth the domain size, or five grid zones.  The pattern
grows rapidly and vertical speeds exceed the gas sound speed after
1.62~orbits.  The rising gas penetrates the overlying shock pattern,
disrupting it by 1.89~orbits.  The secondary instability resembles the
Shaviv type I instability in that overturning motions lead to the
growth of low-density chimneys of rising material, but differs in that
the horizontal wavelength is six times shorter.  The secondary
instability might be a form of the two-dimensional type~I instability
modified by the shear $\partial v_x/\partial y$ and if that is the
case, may also require a special lower boundary condition.

\subsection{Dependence on Magnetic Field\label{sec:shockb}}

\begin{figure}
\epsscale{0.8}
\plotone{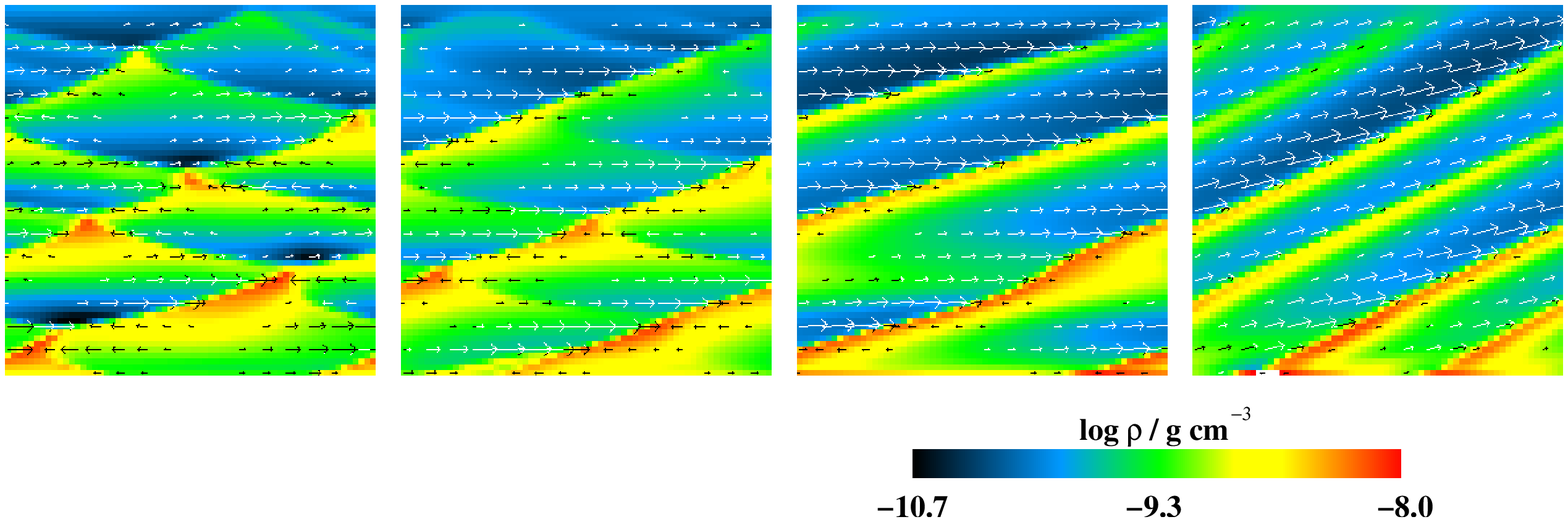}
\caption{Shock trains on magnetic fields of different orientations.
The surface-layer calculations have magnetic pressure 25\% of the
radiation pressure at domain center.  The fields are initially
inclined 0, 1, 3 and $12^\circ$ (left to right).  Results are shown at
1.8~orbits after the shock merger phase has ended, except at right
where the time is 1.2~orbits.  Density is indicated by colors on a
shared logarithmic scale, velocity by arrows.
\label{fig:shockangleb}}
\end{figure}

The magnetic field orientation affects the growth rate and propagation
direction of the fastest photon bubble mode (section~\ref{sec:linear})
and therefore how quickly shocks develop and at what angle they first
appear.  The orientation of the field is varied in a series of
calculations with $64^2$ zones that are otherwise identical to the
fiducial surface-layer run of section~\ref{sec:shockfiducial}.  On
horizontal fields, modes propagating to left and right grow at equal
rates, leading to a pattern of crossed shocks shown in
figure~\ref{fig:shockangleb}.  On fields inclined three degrees or
more from horizontal, the shock pattern is dominated by fronts of one
inclination, propagating downhill along the field.  Each calculation
passes through the same three phases as the fiducial calculation,
ending with a single shock in the domain width.  In the calculation
with horizontal fields, both the left and right-moving fronts have
horizontal separation equal to the domain width.  On fields inclined
12~and 24~degrees, during the shock merging phase, gas in low-density
regions near the fixed lower boundary is accelerated upward by the
radiation force, leaving behind extremely low densities.  The
corresponding large fluxes lead to bending of the field and disruption
of the pattern.

\begin{figure}
\epsscale{0.75}
\plotone{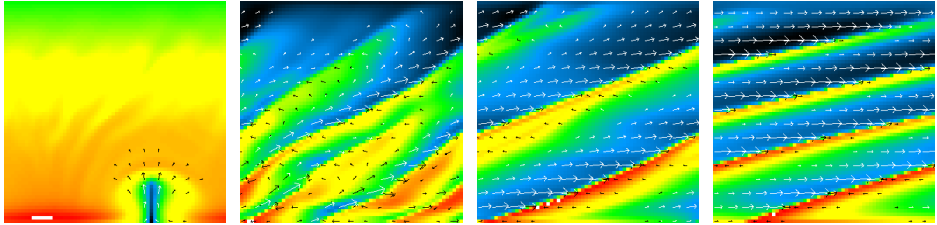}
\caption{Shock trains on magnetic fields of different strengths.  All
the calculations treat the disk surface layers and have fields
inclined $3^\circ$.  The magnetic pressures are 0.1, 1, 10 and 100\%
of the domain-center radiation pressure (left to right).  Density
(colors) and velocity (arrows) are shown at different times in the
different calculations: 0.8, 1.0, 1.4 and 2.2~orbits (left to right).
The color scale is logarithmic between the largest and smallest
density in each panel and the longest arrows correspond to speeds
$2.6\times 10^7$, $1.0\times 10^7$, $2.7\times 10^7$ and $2.9\times
10^7$ cm s$^{-1}$.  The shock pattern is disrupted by the overturning
instability in the case with weakest magnetic fields and by field
bending in the two middle cases.  The pattern is long-lasting in the
strong-field case.
\label{fig:shockpmag}}
\end{figure}

The strength of the magnetic field affects the shocks through field
line bending.  A series of surface-layer calculations was made with
fields having different strengths but a common inclination of
3~degrees.  The initial magnetic pressures are 0.1, 1, 10 and 100\% of
the radiation pressure at domain center or 0.254 to 254 times the
domain-center gas pressure.  A snapshot from each calculation is shown
in figure~\ref{fig:shockpmag}.  The results are described below and
may be compared with the fiducial calculation, where the magnetic
pressure is 25\% of the domain-center radiation pressure.
\begin{enumerate}
\item In the weakest-field case, the first instabilities reaching
  non-linear amplitudes are photon bubbles in the upper two-thirds of
  the domain and the Shaviv type~I overturn instability in the lower
  one-third.  The magnetic fields are bent by gas pressure gradient
  and radiation forces when the disturbances reach non-linear
  amplitudes.  At 0.6~orbits, the largest density variations due to
  photon bubbles are 5\%, those due to the overturn instability are
  9\% and the field inclination ranges seven degrees either side of
  its starting value.  At 0.8~orbits the speeds due to the photon
  bubbles approach the gas sound speed.  The evolution after
  0.8~orbits is dominated by a vertical chimney of rising, low-density
  gas that develops from the overturn instability and pushes aside the
  magnetic fields and weak shock train.
\item In the case with magnetic pressure 1\% of the domain-center
  radiation pressure, field lines are first bent by more than five
  degrees at 0.7~orbits, when the largest ram pressure, $1\,300$ dyn
  cm$^{-2}$, is still much less than the magnetic and gas pressures at
  the same location, $37\,000$ and $33\,000$ dyn cm$^{-2}$,
  respectively.  The forces bending the field lines are roughly equal
  parts due to the flux perturbations and gas pressure gradients from
  the density variations in the photon bubbles.  Gas collects near
  shock fronts at inflection points in the field lines, further
  increasing the density contrast which reaches a factor five at one
  orbit.  The low-density gas is then driven up parallel to the shock
  fronts by the radiation force and regions near the bottom boundary
  are evacuated.  The calculation ends due to low densities after
  1.13~orbits when the horizontally-averaged radiation flux is
  2.6~times greater than initially.
\item In the case with pressure ratio 10\%, field lines first bend
  significantly during the shock merger phase.  Mergers end at
  1.4~orbits with the front spacing equal to the domain width.  The
  largest ram pressure in the domain is about equal to the magnetic
  pressure and the field is bent by up to 8~degrees.  At 1.48~orbits
  the horizontally-averaged flux is 4.8~times the initial value.  The
  field is lifted away from the lower boundary at one location by
  radiation forces, leaving behind very low densities so that
  timesteps are short and further calculation is impractical.
\item In the strongest-field case, the field lines remain within
  0.7~degrees of their initial orientation for the six-orbit duration
  of the run.  The ram pressure of the gas is everywhere less than 5\%
  of the magnetic pressure.  From 1.6~orbits on, the horizontal
  distance between fronts is equal to the domain width and the shock
  pattern varies slowly due to the vertical gradient in the
  propagation speed.  Overdense gas moving down field lines near the
  lower boundary rises again only with difficulty, because the
  vertical velocity is fixed at zero on the boundary.  Material
  gradually accumulates near the bottom and the calculation is ended
  after the horizontally-averaged density at domain center falls below
  half the initial value.
\end{enumerate}
In summary, photon bubbles lead to shock trains over the whole range
of field strengths explored.  The shock amplitude is limited by the
growth of the overturning instability in the case with magnetic
pressure less than gas pressure, by the strength of the magnetic field
in the two intermediate cases and by the size of the domain in the
calculation with the strongest fields.

\subsection{Low Optical Depth\label{sec:shocktau}}

Photon bubbles saturate at low amplitudes in a calculation centered on
the disk photosphere at $z=1.12 H$.  The square domain extends from
1.07 to $1.17 H$ and is filled with an initial condition generated
using the density, temperature and flux at the photosphere in the
full-disk-thickness calculations.  The magnetic field is horizontal
with pressure 10\% of the midplane radiation pressure, and random 1\%
density perturbations are applied throughout the domain.  The grid
resolution is $32^2$.  Photon bubbles grow until gas moves back and
forth along the field lines with velocity amplitude twice the
isothermal gas sound speed and the photosphere is crossed by weak
shocks.  The speeds are greatest below the photosphere but the kinetic
pressure is everywhere less than 1\% of the magnetic pressure so that
field lines remain almost straight.  Eddington factors depart slightly
from one-third in regions of low density near the photosphere and the
results may depend on the angular variation of the radiation
intensity.  More detailed calculations in this regime may be useful to
test the accuracy of the relationship between the radiation energy
gradient and the anisotropy assumed in our flux-limited diffusion
calculation.  However the saturation at low amplitude is a product of
the artificial lower boundary in the surface-layer calculation.  In
the case spanning the disk thickness shown in figure~\ref{fig:3up},
the photosphere is disrupted after shock trains develop deeper in the
disk.

\subsection{Full Disk Thickness\label{sec:shockfull}}

In this section we explore the growth of shocks and the effects on
cooling in the calculation with magnetic fields shown in
figure~\ref{fig:3up}.  The calculation extends through the thickness
of the disk from one photosphere to the other.  Photon bubbles grow
from initial 1\% random density perturbations applied within $1.01 H$
of the midplane.  The fastest linear mode between heights $0.65$ and
$1.05 H$ has 14 wavelengths in the domain width and 5 in the height of
the region and grows at $3.73 \Omega$.  The fastest mode grows slower,
has longer wavelength and propagates further from the horizontal than
in the calculation described in section~\ref{sec:pbifast}, due to the
lower resolution here.  Density disturbances exceed 10\% after
0.18~orbits and the photon bubbles develop into trains of shocks.  The
shock fronts are curved because the wavevector of the fastest-growing
linear mode varies with height, being more nearly parallel to the
magnetic field in the disk interior.  The relationship between flux
and overdensity at $z=H$ is consistent with the improved analytic
shock train solution until 0.48~orbits.  The radiation force due to
the increased flux then ejects the surface layers above $z=H$.  By
0.7~orbits 2\% of the total mass is lost, while the flux increases
above that expected from the analytic solution.  Magnetic field lines
are bent increasingly by the ram pressure of the gas, and near the
strongest shocks depart more than $10^\circ$ from their initial
orientation after 0.6~orbits.  Fluxes locally exceed 21~times the
initial photospheric flux $F_1$ in the low-density channels between
the shocks at 0.8~orbits and the radiation force drives the fronts
toward the vertical.  A further 0.7\% of the initial mass is ejected
through the channels by 1.5~orbits, when the total radiation energy
has fallen by 94\% and the calculation is ended.  The declining fluxes
in the later stages are not sufficient to support the overdense gas in
the surface layers and the dense material collapses toward the
midplane.

The timestep in the calculation shown in figure~\ref{fig:3up} is
100~times the diffusion step determined by the lowest densities.  The
effects of the timestep are checked using a version with steps ten
times shorter that is run for one orbit.  Linear photon bubble growth
rates are similar and the same fastest mode grows at $3.77 \Omega$.
Owing to a more accurate rise in the radiation energy outside the
photosphere in response to increased fluxes inside, a smaller fraction
of the surface layers is ejected and the mass decreases by only 1.2\%
up to 0.7~orbits.  The mean cooling rate between 0.7 and 1.0~orbits is
slightly faster than in the calculation with longer timesteps, at
5.9~times the rate in the calculation with diffusion alone.  Overall,
the results depend little on the timestep.  However the effects of the
photon bubbles are underestimated in both calculations because the
longest wavelength of rapid linear growth, $2\pi c_i^2/g$, is
marginally resolved in the surface layers.  Faster linear growth at
short wavelengths is indicated by the WKB analysis and measured in the
higher-resolution calculation described in section~\ref{sec:pbifast}.

\begin{figure}
\epsscale{0.59}
\plotone{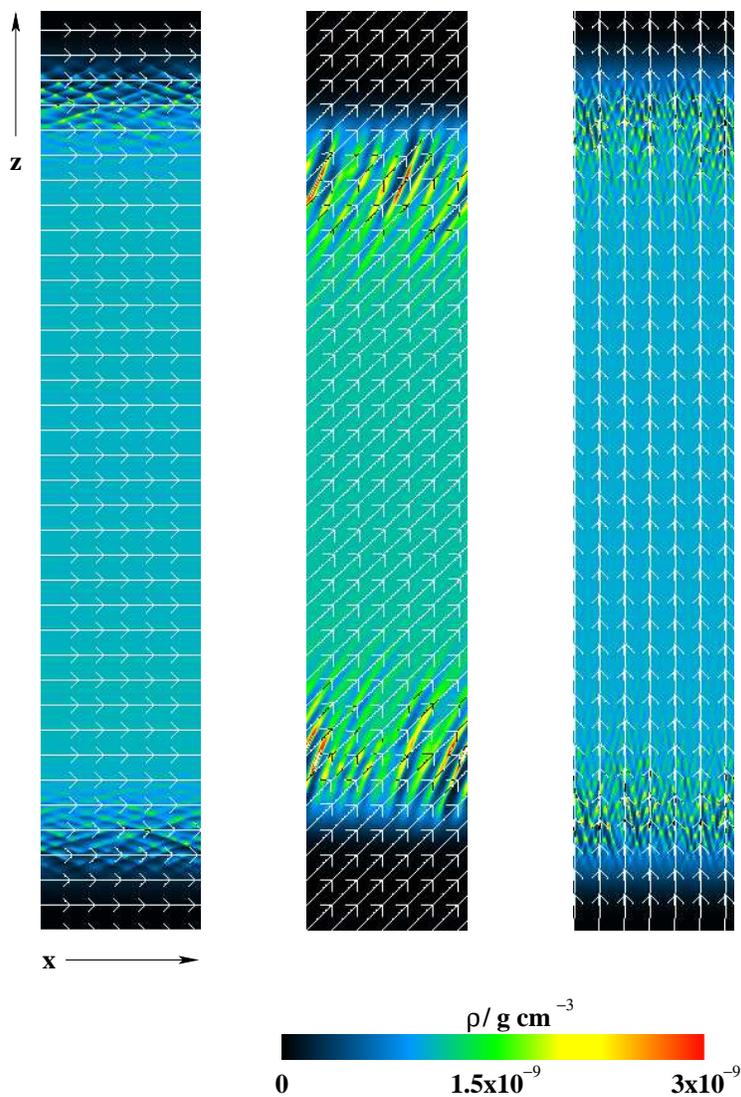}
\caption{Photon bubbles in calculations spanning the disk thickness,
with magnetic fields initially horizontal (left) and inclined
$45^\circ$ (center) and $87^\circ$ (right).  Other parameters are as
in figure~\ref{fig:3up}.  At the times shown, the horizontal
variations in density first exceed a factor ten.  The times are
1.22~orbits (left), 0.72~orbits (center) and 0.33~orbits (right).
Densities are indicated by colors on a common linear scale, magnetic
fields by arrows with the longest corresponding to 5200~Gauss.  The
largest local radiation fluxes are found in the $45^\circ$ case, where
shocks tilted to one side dominate and low-density channels extend
deep into the disk.
\label{fig:density10}}
\end{figure}

The dependence on the orientation of the magnetic field is examined
with two additional calculations extending through the disk thickness.
The only parameter differing from figure~\ref{fig:3up} is the initial
inclination of the field.  The field is horizontal in one case and
inclined 45~degrees in the other.  Results at the times when
horizontal density variations first exceed a factor ten are shown in
figure~\ref{fig:density10} with a corresponding view of the case from
figure~\ref{fig:3up} having the field inclined 87~degrees.  The times
shown are 1.22, 0.72 and 0.33~orbits, respectively.  As expected from
the linear analysis, the development is fastest on near-vertical
fields.  The effects of the photon bubbles are also large in the case
with the intermediate inclination, where modes with fronts tilted to
the same side as the field grow much faster than those with fronts
tilted the other way and long, unobstructed channels of low density
are formed.  The largest radiation flux is 30\% greater than at the
same density contrast in the case from figure~\ref{fig:3up}.  The
shock spacing and amplitude and the gas speeds continue to grow until
at 0.94~orbits the ram pressure exceeds the magnetic pressure at some
locations and the field lines are bent.  Gas collects near the
inflection points in the field and slides down the field lines under
its own weight, disrupting the shock pattern in the interior but
leaving a regular train of fronts in the surface layers.  The
calculation is ended at 1.2~orbits.  In the case with horizontal
field, shock mergers by 1.5~orbits increase the horizontal spacing of
the fronts to $0.1 H$ at height $z=H$.  The maximum fluxes at this
time are twice the initial photospheric flux.  The radiation force
then drives off the outer layers where no perturbations were placed
initially.  By the end of the calculation at 1.8~orbits, 13\% of the
mass is lost through the top and bottom boundaries.

\section{SUMMARY AND CONCLUSIONS\label{sec:conc}}

We study the cooling of radiation-pressure dominated accretion disks
due to convection and the Shaviv type~I and photon bubble
instabilities, using 2-D and 3-D radiation-MHD calculations of small
patches of disk.  The background conditions are simplified by
neglecting differential rotation, so the turbulence and heating
resulting from MRI are absent.  The initial states are chosen from a
Shakura-Sunyaev model, but no effective viscosity is applied during
the calculations.  The cooling caused by the instabilities is compared
against the 1-D vertical diffusion that is assumed in the
Shakura-Sunyaev picture.
\begin{enumerate}
\item In the absence of magnetic fields, convective instability grows
at about the orbital frequency $\Omega$ and cools the disk 60\% faster
than 1-D diffusion, consistent with previous studies by \cite{pk00}
and \cite{ak01}.
\item The Shaviv type~I global instability is not observed in
calculations extending through the disk thickness, but grows quickly
in calculations with a closed lower boundary of fixed temperature.
The wavelengths and fastest growth rates agree approximately with
those in the linear analysis by \cite{sh01} despite different
assumptions about the background state.  Disturbances reaching large
amplitudes become narrow chimneys of rising low-density gas separated
by denser, sinking material.  The Shaviv type~I instability is
unlikely to have strong effects in radiation-dominated disks but may
be important in radiation-supported atmospheres with distinct lower
boundaries.
\item Photon bubbles with wavelengths shorter than the gas pressure
scale height $c_i^2/g$ grow faster than the orbital frequency when
magnetic pressure exceeds gas pressure, in detailed agreement with a
linear WKB analysis.  Growth approaches the asymptotic rate $g/(2
c_i)$ expected from the linear analysis, when the field is nearly
vertical.  The disturbances develop into trains of propagating shocks
similar to those predicted by \cite{be01}.  The shocks grow in
strength and increase in separation over time, while radiation escapes
through the gaps between, cooling the flow in the case we considered
five times faster than assumed in the Shakura-Sunyaev model.
\end{enumerate}
Unlike convection, which is driven by an entropy gradient and can be
absent if heating is concentrated in the surface layers, photon
bubbles are driven by the radiation flux.  The photon bubble shoulder
wavenumber $g/c_i^2$ and the fastest growth rate $g/(2 c_i)$ at the
disk surface are determined by the temperature and height of the
photosphere.  They are independent of the details of the internal
structure and independent of the field strength, provided the magnetic
pressure is greater than the gas pressure.  Instability is present if
the ratio of the flux to the radiation energy density exceeds
approximately the isothermal gas sound speed $c_i$ in optically-thick
regions having radiation and magnetic pressures greater than the gas
pressure \citep{bs03}.  Since the ratio of flux to energy density must
approach $c$ at the photosphere and $c > c_r > c_i$, the criterion for
linear instability is likely to be satisfied in the outer layers.

A major issue to be resolved is the amplitudes reached by photon
bubbles in turbulence driven by the MRI, where structures are
destroyed on the eddy turnover timescale of about an orbit.  Photon
bubble growth rates depend on the orientation of the magnetic fields.
The fields in MRI turbulence have a small mean vertical component in
3-D shearing-box MHD calculations of patches of accretion disk
neglecting radiation diffusion \citep{bn95,sh96,ms00}, and the median
angle between magnetic field and midplane is $10^\circ$ in a
calculation including diffusion \citep{tu04}.  On fields with similar
inclinations in the calculations described in
section~\ref{sec:linear}, photon bubbles grow faster than $2 \Omega$,
corresponding to an increase in amplitude by a factor $3\times 10^5$
per orbit.  The fastest modes in our numerical calculations grow on
nearly vertical magnetic fields at $5 \Omega$, increasing in amplitude
by a factor $4\times 10^{13}$ per orbit.  Given time enough to grow,
small disturbances develop into trains of shocks that increase in
strength and wavelength until the magnetic fields buckle.  The
dependence of the linear growth rate on field angle could lead to
effects varying in time and space according to the field structures
found in the turbulence.

Fast photon bubble growth consistent with the asymptotic regime of the
linear analysis is found only in calculations with grid spacing less
than one-tenth the shoulder wavelength $2\pi c_i^2/g$ and timestep
shorter than the diffusion step.  Accurately following the growth of
photon bubbles throughout the disk thickness for the parameters we
chose required a grid of about a thousand zones in the vertical
direction.  The grid spacing needed in a Shakura-Sunyaev disk model is
about $H p_1/P_c$, where $p_1$ is the gas pressure at the photosphere
and $P_c$ is the radiation pressure at the midplane.  Future advances
in computer technology will enable the study of photon bubbles
together with the MRI using 3-D radiation-MHD calculations on such
grids.

The analytic and numerical results indicate that photon bubbles
develop into regions of alternating high and low density much smaller
than the disk thickness.  Radiation escapes faster through the porous
flow than through the hydrostatic structure of the same surface
density and the resulting cooling may have fundamental effects on the
overall thermal balance.  Inhomogeneities within a few Thomson depths
of the photosphere can modify the emitted thermal spectrum
\citep{db04} and the reflected X-ray spectrum \citep{bt04} while
sufficiently rapid cooling could lead to super-Eddington luminosities
from geometrically thin accretion disks \citep{be02}.

%%%%%%%%%%%%%%%%%%%%%%%%%%%%%%%%%%%%%%%%%%%%%%%%%%%%%%%%%%%%%%%%%%%%%%%%%%%%%%%
\begin{acknowledgments}
This work was supported by the National Science Foundation under grant
AST-0307657, by NASA under grant NAG5-12035 in the Astrophysical
Theory Program and by the National Research Council through a
fellowship to N. J. T. at the Jet Propulsion Laboratory, California
Institute of Technology.
\end{acknowledgments}

%%%%%%%%%%%%%%%%%%%%%%%%%%%%%%%%%%%%%%%%%%%%%%%%%%%%%%%%%%%%%%%%%%%%%%%%%%%%%%%
\appendix
\section{How the Photon Bubble Instability Works}

\newcommand{\be}{  \begin{eqnarray} }
\newcommand{\ee}{  \end{eqnarray} }
\newcommand{\cross}{\times}

\def\crad{c_{\rm r}}
\def\cgas{c_{\rm g}}
\def\vaz{v_{{\rm A}z}}
\def\vaphi{v_{{\rm A}\phi}}
\def\vph{v_{\rm ph}}
\def\msun{{\rm M}_\odot}
\def\lsun{{\rm L}_\odot}
\def\ledd{{\rm L}_{\rm Edd}}
\def\kappaj{\kappa_{\rm J}}
\def\kappap{\kappa_{\rm P}}
\def\kappat{\kappa_{\rm T}}
\def\kappaf{\kappa_{\rm F}}
\def\sigmat{\sigma_{\rm T}}
\def\kb{k_{\rm B}}
\def\omegaa{\omega_{\rm a}}
\def\omegath{\omega_{\rm th}}
\def\kf{{\bf k}\cdot{\bf F}}
\def\thetarho{\Theta_\rho}
\def\thetatg{\Theta_{T{\rm g}}}
\def\thetatr{\Theta_{T{\rm r}}}
\def\thetat{\Theta_T}
\def\spose#1{\hbox to 0pt{#1\hss}}
\def\lta{\mathrel{\spose{\lower 3pt\hbox{$\mathchar"218$}}
     \raise 2.0pt\hbox{$\mathchar"13C$}}}
\def\gta{\mathrel{\spose{\lower 3pt\hbox{$\mathchar"218$}}
     \raise 2.0pt\hbox{$\mathchar"13E$}}}
\font\syvec=cmbsy10                        %for boldface nabla
\font\gkvec=cmmib10                         %for boldface lowercase Greek
\def\bnabla{\hbox{{\syvec\char114}}}       %bold face nabla
\def\balpha{\hbox{{\gkvec\char11}}}        %bold face alpha
\def\bbeta{\hbox{{\gkvec\char12}}}         %bold face beta
\def\bgamma{\hbox{{\gkvec\char13}}}        %bold face gamma
\def\bdelta{\hbox{{\gkvec\char14}}}        %bold face delta
\def\bepsilon{\hbox{{\gkvec\char15}}}      %bold face epsilon
\def\bzeta{\hbox{{\gkvec\char16}}}         %bold face zeta
\def\boldeta{\hbox{{\gkvec\char17}}}       %bold face eta
\def\btheta{\hbox{{\gkvec\char18}}}        %bold face theta
\def\biota{\hbox{{\gkvec\char19}}}         %bold face iota
\def\bkappa{\hbox{{\gkvec\char20}}}        %bold face kappa
\def\blambda{\hbox{{\gkvec\char21}}}       %bold face lambda
\def\bmu{\hbox{{\gkvec\char22}}}           %bold face mu
\def\bnu{\hbox{{\gkvec\char23}}}           %bold face nu
\def\bxi{\hbox{{\gkvec\char24}}}           %bold face xi
\def\bpi{\hbox{{\gkvec\char25}}}           %bold face pi
\def\brho{\hbox{{\gkvec\char26}}}          %bold face rho
\def\bsigma{\hbox{{\gkvec\char27}}}        %bold face sigma
\def\btau{\hbox{{\gkvec\char28}}}          %bold face tau
\def\bupsilon{\hbox{{\gkvec\char29}}}      %bold face upsilon
\def\bphi{\hbox{{\gkvec\char30}}}          %bold face phi
\def\bchi{\hbox{{\gkvec\char31}}}          %bold face chi
\def\bpsi{\hbox{{\gkvec\char32}}}          %bold face psi
\def\bomega{\hbox{{\gkvec\char33}}}        %bold face omega

As discussed at length in \cite{bs03}, there are two basic physical
mechanisms for the radiative amplification of traveling acoustic
waves.  The first relies on an interaction between the background
radiative flux ${\bf F}$ and fluctuations in the flux mean (Rosseland)
opacity $\kappa_{\rm F}$ in the wave, and exists even in the absence
of magnetic fields.  This mechanism is irrelevant for the conditions
we consider in this paper, where the flux mean opacity is dominated by
Thomson scattering, and is therefore constant.  The second mechanism
originates from a breakdown in cancellation of background radiation
pressure, gas pressure, and gravitational forces on perturbed fluid
elements.  This can only occur for waves which are not purely
longitudinal in character, which for acoustic waves requires a
background magnetic field.  It turns out that the force responsible
for this second driving mechanism can be expressed entirely in terms
of the flux perturbation that is perpendicular to the wave vector of
the wave, and we show how to do this here.  We used this fact in our
illustration of the physical mechanism of the instability in
figure~\ref{fig:mechanism}.
 
The linearized equation of motion of a fluid element in a short
wavelength magneto\-acoustic wave may be written as (equations 74, 75,
and 100 of \cite{bs03})
\begin{eqnarray}
\rho{\partial^2{\bxi}\over\partial t^2}&=&\rho\left[
-({\bf k}\cdot{\bf v}_{\rm A})^2{\bxi}
+({\bf k}\cdot{\bxi})({\bf k}\cdot{\bf v}_{\rm A}){\bf v}_{\rm A}
+({\bf k}\times{\bf v}_{\rm A})\cdot({\bf v}_{\rm A}\times{\bxi}){\bf k}
-{\bf k}\cgas^2({\bf k}\cdot{\bxi})\right]\nonumber\\
&+&i({\bf k}\cdot{\bxi}){\bf k}{\kappaf\rho\over c}\left\{
{4E\over3}\left({\omega\over k^2}\right)+{4Ec\omegaa(\gamma-1)^2\over\kappaf
\rho\omega}-{1\over k^2}({\bf k}\cdot{\bf F})\left[\thetarho+(\gamma-1)
\thetatg\right]\right\}\nonumber\\
&-&i\left[{\kappaf\rho\over k^2c}{\bf k}\left({\bf k}\cdot{\bf F}\right)\left(
{\bf k}\cdot{\bxi}\right)-{\bf k}{\bxi}\cdot{\bnabla}p+
\rho{\bf g}\left({\bf k}\cdot{\bxi}\right)\right]+
{\cal O}(k^0)|{\bxi}|.
\label{mhdmomhighk}
\end{eqnarray}
Here ${\bxi}$ is the Lagrangian displacement vector of the fluid
element under consideration, ${\bf k}$ is the wavevector of the wave
and $k$ is its magnitude, $\omega$ is the angular frequency of the
wave, ${\bf v}_{\rm A}$ is the Alfv\'en velocity, $\cgas$ is the
adiabatic sound speed in the gas, $p$ is the gas pressure, $\rho$ is
the density, $\gamma$ is the adiabatic index of the gas, $E$ is the
radiation energy density, $c$ is the speed of light, $\omega_{\rm a}$
is a characteristic angular frequency associated with true absorption
opacity (equation 30 of \cite{bs03}), ${\bf g}$ is the acceleration
due to gravity, $\Theta_{\rho}$ is the logarithmic derivative of
$\kappa_{\rm F}$ with respect to density, and $\Theta_{T{\rm g}}$ is
the logarithmic derivative of $\kappa_{\rm F}$ with respect to the gas
temperature.

The first square bracket term on the right hand side of equation
(\ref{mhdmomhighk}) represents the magnetic pressure, magnetic
tension, and gas pressure terms that provide the basic restoring
forces that support the wave.  Radiation pressure does not contribute
here because of rapid photon diffusion at these short wavelengths.
The three terms in curly braces represent the effects of Silk damping,
damping of temperature differences between the gas and radiation due
to absorption and emission, and damping (or driving) due to the
interaction between the background radiative flux and opacity
fluctuations in the wave.

The last term in square brackets on the right hand side of equation
(\ref{mhdmomhighk}) is the term of greatest interest here.  It
represents the possible unstable driving of magneto\-acoustic waves
due to the interplay of background radiation pressure gradient, gas
pressure gradient, and gravitational forces.  Using the equation of
hydrostatic equilibrium in the background,
\begin{equation}
0=-{\bnabla} p+{\kappa_{\rm F}\rho\over c}{\bf F}+\rho{\bf g},
\label{hydrostatic}
\end{equation}
\cite{bs03} eliminated the gas pressure gradient to show that this
driving term could be written as
\begin{equation}
{\kappaf\rho\over k^2c}{\bf k}\left({\bf k}\cdot{\bf F}\right)\left(
{\bf k}\cdot{\bxi}\right)-{\bf k}{\bxi}\cdot{\bnabla}p+\rho{\bf g}\left(
{\bf k}\cdot{\bxi}\right)=
{\kappaf\rho\over c}{\bf k}
\left[(\hat{\bf k}\cdot{\bf F})(\hat{\bf k}\cdot{\bxi})
-({\bxi}\cdot{\bf F})\right]-\rho{\bxi}\times({\bf k}\times{\bf g}).
\end{equation}
The last term, involving the gravitational acceleration, is
perpendicular to the fluid displacement.  It therefore does no work,
and provides neither damping or driving.  The first term involving the
radiative flux ${\bf F}$ represents a potential driving force based on
an interaction between ${\bf F}$ and density fluctuations in the wave.
The second term involving ${\bf F}$ is a driving force arising from a
change in the background radiation pressure along a fluid
displacement.  The sum of these two terms is proportional to the
gradient of the Lagrangian change in radiation pressure \citep{bs03},
and it is important to note that they cancel precisely for any
longitudinal wave.  Because magneto\-acoustic waves need not be
longitudinal, such waves can be driven unstable by these terms.

It turns out that it is also possible to write these driving terms in
another way that is also physically intuitive.  If we use hydrostatic
equilibrium to eliminate ${\bf g}$ rather than the gas pressure
gradient, then
\begin{equation}
-i\left\{{\kappaf\rho\over k^2c}{\bf k}\left({\bf k}\cdot{\bf F}\right)\left(
{\bf k}\cdot{\bxi}\right)-{\bf k}{\bxi}\cdot{\bnabla}p+\rho{\bf g}\left(
{\bf k}\cdot{\bxi}\right)\right\}
={\kappa_{\rm F}\rho\over c}\delta{\bf F}_{\perp,
\Theta=0}+i{\bxi}\times({\bf k}\times{\bnabla}p).
\end{equation}
Once again, the second term on the right hand side is perpendicular to the
fluid displacements, and does no work.  In the first term,
\begin{equation}
\delta {\bf F}_{\perp,\Theta=0}\equiv\left[\hat{\bf k}(\hat{\bf k}\cdot{\bf F})
-{\bf F}\right]{\delta\rho\over\rho}
\end{equation}
is the component of the flux perturbation that is perpendicular to the
wave vector ${\bf k}$, ignoring opacity fluctuations in the wave.
Physically, a compressive plane wave will naturally produce positive
(negative) flux perturbations through the rarefied (compressed)
regions due to the increased (decreased) transparency of the medium
(figure~\ref{fig:mechanism}).  Because surfaces of constant density in
the wave follow surfaces of constant phase, these flux perturbations
will be perpendicular to the wave vector.  In a wave that is not
purely longitudinal, these flux perturbations will have nonvanishing
projections onto the fluid velocity in the wave, and can therefore
cause damping or driving of the wave.

The discussion so far was for the case where the gas and radiation
temperatures in the medium are allowed to depart from exact equality.
In all the calculations shown in this paper, photon absorption and
emission are so fast that the two temperatures are generally locked
together.  In this case, the equation of motion of a perturbed fluid
element is (equations 82, 84, and 106 of \cite{bs03})
\begin{eqnarray}
\rho{\partial^2{\bxi}\over\partial t^2}&=&\rho\left[
-({\bf k}\cdot{\bf v}_{\rm A})^2{\bxi}
+({\bf k}\cdot{\bxi})({\bf k}\cdot{\bf v}_{\rm A}){\bf v}_{\rm A}
+({\bf k}\times{\bf v}_{\rm A})\cdot({\bf v}_{\rm A}\times{\bxi}){\bf k}
-{\bf k}c_{\rm i}^2({\bf k}\cdot{\bxi})\right]\nonumber\\
&+&i({\bf k}\cdot{\bxi}){\bf k}{\kappaf\rho\over c}\left\{
\left(1+{3p\over4E}\right)
\left[\left({4E\over3}+p\right){\omega\over k^2}-{1\over k^2}
({\bf k}\cdot{\bf F})\thetarho\right]\right\}\nonumber\\
&-&i\left[\left(1+{3p\over4E}\right){\kappaf\rho\over k^2c}{\bf k}({\bf k}\cdot
{\bf F})({\bf k}\cdot{\bxi})-{\bf k}c_{\rm i}^2{\bxi}\cdot{\bnabla}\rho
+\rho{\bf g}({\bf k}\cdot{\bxi})\right]+{\cal O}(k^0)|{\bxi}|.
\label{mhdmomhighklock}
\end{eqnarray}
Here the adiabatic sound speed in the gas in the first square brackets
term on the right hand side has been replaced by the isothermal sound
speed $c_{\rm i}$, because gas temperature fluctuations in the wave
are smoothed out by the rapid radiative diffusion.  Similarly, only
density-induced opacity fluctuations are relevant in the
damping/driving term in the term in braces.

Once again, it is the last square bracket term that is most relevant
here.  Hydrostatic equilibrium and radiative diffusion in the
background implies
\begin{equation}
\rho{\bf g}=c_{\rm i}^2{\bnabla}\rho-\left(1+{3p\over4E}\right)
{\kappaf\rho\over c}{\bf F}.
\end{equation}
Using this to eliminate the acceleration due to gravity, we find
\begin{eqnarray}
& &-i\left[\left(1+{3p\over4E}\right){\kappaf\rho\over k^2c}{\bf k}({\bf k}
\cdot{\bf F})({\bf k}\cdot{\bxi})-{\bf k}c_{\rm i}^2{\bxi}\cdot{\bnabla}\rho
+\rho{\bf g}({\bf k}\cdot{\bxi})\right]\nonumber\\
&=&
\left(1+{3p\over4E}\right){\kappa_{\rm F}\rho\over c}\delta{\bf F}_{\perp,
\Theta=0}+ic_{\rm i}^2{\bxi}\times({\bf k}\times{\bnabla}\rho).
\end{eqnarray}
Once again, the last term involving the density gradient does no work.
In the absence of opacity fluctuations, the driving can be seen to be
due entirely to the component of the flux perturbation that is
perpendicular to the wave vector.

%%%%%%%%%%%%%%%%%%%%%%%%%%%%%%%%%%%%%%%%%%%%%%%%%%%%%%%%%%%%%%%%%%%%%%%%%%%%%%%

\end{document}